\documentclass[showpacs,aps,prl,twocolumn,superscriptaddress]{revtex4-1}

\usepackage[english]{babel}
\usepackage{graphicx}
\usepackage{hyperref}
\usepackage{epsfig}
\usepackage{color}
\usepackage{psfrag}
\usepackage{float}
\usepackage{tikz}
\usepackage{amsmath}
\usepackage{amssymb}
\usepackage{float}

\newcommand{\ket}[1] {\left|#1\right\rangle}

\newcommand{\Figref}[1]{Fig.~\ref{#1}}
\newcommand{\Eqref}[1]{Eq.~(\ref{#1})}
\newcommand{\eref}[1]{(\ref{#1})} 

\begin{document}
%
\title{Photon-Emission Statistics induced by Electron Tunnelling in Plasmonic Nanojunctions}
%

%
\author{R. Avriller}
\affiliation{Univ. Bordeaux, CNRS, LOMA, UMR 5798, F-33405 Talence, France}
\author{Q. Schaeverbeke}
\affiliation{Univ. Bordeaux, CNRS, LOMA, UMR 5798, F-33405 Talence, France}
\affiliation{Donostia International Physics Center (DIPC), E-20018 Donostia-San Sebasti\'{a}n, Spain}
\author{T. Frederiksen}
\affiliation{Donostia International Physics Center (DIPC), E-20018 Donostia-San Sebasti\'{a}n, Spain}
\affiliation{Ikerbasque, Basque Foundation for Science, E-48013, Bilbao, Spain}
\author{F. Pistolesi}
\affiliation{Univ. Bordeaux, CNRS, LOMA, UMR 5798, F-33405 Talence, France}
\date{\today}

\begin{abstract}
We investigate the statistics of photons emitted by tunneling electrons in a single electronic level plasmonic nanojunction.  
We compute the waiting-time distribution of successive emitted photons $w(\tau)$.
When the cavity damping rate $\kappa$ is larger than the electronic 
tunneling rate $\Gamma$, we show that in the photon-antibunching regime, 
$w(\tau)$ indicates that the average delay-time between two successive photon 
emission events is given by $1/\Gamma$.
This is in contrast with the usually considered second-order correlation function of emitted photons, $g^{(2)}(\tau)$, which displays the single time scale $1/\kappa$.
Our analysis shows a relevant example for which $w(\tau)$ gives  independent information 
on the photon-emission statistics with respect to $g^{(2)}(\tau)$, leading to a physical insight on the 
problem. 
We discuss how this information can be extracted from experiments even in presence of a 
non-perfect photon detection yield.
\end{abstract}

\maketitle

%
The correlation functions of the electromagnetic field are known to contain a rich amount of information about the intrinsic quantum nature of the electromagnetic field, as well as of the sources at the origin of photon emission \cite{PhysRev.130.2529}.
The second-order correlation function (SCF) of the electromagnetic field $g^{(2)}(\tau)$, is of particular interest to investigate the statistics of photons emitted by fluorescent atoms or molecules \cite{reynaud1983fluorescence,carmichael1989photoelectron,cohen1979atoms}. 
It was shown that for a single-photon source, $g^{(2)}(\tau)$ vanishes at short times, a phenomenon known as photon antibunching \cite{reynaud1983fluorescence}.
In the case that the emitter is a single atom or a molecule, photon antibunching is interpreted as arising from the wave-packet projection assumption of quantum mechanics  \cite{carmichael1989photoelectron,cohen1979atoms} : after the first single-photon is emitted, the atom is projected back to its ground state.
The emission of the next single-photon will then necessitate a finite delay-time during which the atom will be excited again, a necessary condition for another spontaneous emission event to occur.
%

%
Photon antibunching which was revealed by measurements of $g^{(2)}(\tau)$ for fluorescent single-molecules deposited on surfaces \cite{verberk2003photon,PhysRevLett.69.1516,huang2016measuring}, is now a cornerstone of molecular spectroscopy.
More recently, the progress in nanotechnologies extended the use of this experimental tool to design a wealth of different single-photon sources made of electrically-driven scanning tunneling microscopes (STM)\cite{zhang2017electrically,merino2015exciton, Atomic-Scale_Dynamics_Probed_Photon_Correlations,
Single_Photon_Emission_Plasmonic_Light_Source_Driven_Coulomb_Blockade,
Photon_Statistics_Electroluminescence_Fano-like_Interference,doppagne2020single},
quantum dots \cite{Electrically_driven_photon_atomic_defect_monolayer_WS2,Yuan102,mizuochi2012electrically,nothaft2012electrically,Electrically_driven_photon_atomic_defect_monolayer_WS2,Yuan102},
nitrogen vacancy centers \cite{mizuochi2012electrically},
single-molecules deposited in molecular crystals \cite{nothaft2012electrically},
and plasmonic nanocavities \cite{gupta2021complex,Electrically_Driven_Single-Photon_Molecular_Chain}.
%
%
\textcolor{black}{The crossover to antibunching in presence of dissipation has also been investigated theoretically for waveguide quantum electrodynamics systems coupled to single atoms \cite{Chen:16,PhysRevA.96.053805}.} 
While most of these works deal with the paradigmatic two-level system model to describe photon antibunching, recent experiments with STM on C$_{60}$ molecular films invoke a Coulomb-blockade mechanism resulting from tip-induced split-off single-level states \cite{Single_Photon_Emission_Plasmonic_Light_Source_Driven_Coulomb_Blockade}.
The actual mechanism at the origin of light-emission in current-driven STM nanojunctions is however still not well understood and might be more complex.
In such systems indeed, current injection is believed to excite the molecule to an electronic excited-state that further decays back to  ground state by emitting a photon.  
Several mechanisms were proposed for describing this molecular excitation, including elastic tunneling of an electron and a hole from the metallic electrodes to the molecule \cite{PhysRevLett.118.127401}, inelastic tunneling of an electron across the junction at the origin of emission of a localized plasmon that is further absorbed by the molecule \cite{PhysRevLett.118.127401,doppagne2018electrofluorochromism}, and a more complex energy-transfer mechanism in which the absorption and emission processes of the localized plasmon by the molecule interfere one with each other \cite{PhysRevLett.119.013901}.
Recently, we theoretically predicted that upon proper tuning of the external gate and electrode potentials, a single electronic level was sufficient to generate electrically-driven single-photon emission \cite{PhysRevLett.123.246601}.
In the same publication we found that $g^{(2)}(\tau)$ relaxes exponentially towards unity on a time scale given by the photon damping-time of the cavity $1/\kappa$ \cite{PhysRevLett.123.246601} and not with the electronic tunneling-time $1/\Gamma$.
This is surprising, since electronic tunneling is the main physical mechanism at the origin of photon emission in the plasmonic nanocavity. 
\textcolor{black}{To our knowledge, in the emerging field of nanoplasmonics there is still no complete understanding of which timescale is actually controlling photon antibunching.
This question is of great experimental relevance to unravel the nature of the light-emission mechanism in current-driven single-photon sources.}
%

%
In this Letter, we investigate theoretically the statistics of photon emission \cite{barsegov2002probing,barsegov2002multidimensional,jakeman1991fluctuations} in a single-level plasmonic nanojunction, going beyond $g^{(2)}(\tau)$.
In particular, we show that the delay-time or waiting-time distribution (WTD) $w(\tau)$ between successive photon-emission events \cite{nienhuis1988photon,Vyas:00}, provides important complementary statistical information to characterize the photon-emission statistics. 
The necessity to carefully discriminate between $w(\tau)$ an $g^{(2)}(\tau)$ is known in molecular fluorescence spectroscopy
\cite{reynaud1983fluorescence,verberk2003photon,lounis2000photon}.
Indeed, standard "start-stop" photon correlators actually measure directly the WTD and not $g^{(2)}(\tau)$ \cite{reynaud1983fluorescence}. 
The difference between both quantities is nevertheless small for measurement times smaller than the average delay-time between photon emission events and for weak photon-detection yields \cite{reynaud1983fluorescence,lounis2000photon}.
The relevance of studying both $w(\tau)$ an $g^{(2)}(\tau)$ was recently noticed and successfully applied to the investigation of the full counting statistics of electronic currents in molecular junctions \cite{brandes2008waiting,PhysRevLett.107.086805,PhysRevLett.108.186806,PhysRevB.92.125435,PhysRevLett.122.247403,PhysRevLett.127.096803}, or of the statistics of photon emission in microwave cavities \cite{PhysRevB.99.085418}.
However, such is not the case for current-induced single-photon sources,
for which most studies do not discriminate clearly between these two quantities. 
\textcolor{black}{We show how the joint calculation of $g^{(2)}(\tau)$ and $w(\tau)$ enables to unveil the timescales involved in the photon-emission process, thus paving the way for using these complementary observables in order to investigate both theoretically and experimentally the various existing mechanisms predicting current-driven photon antibunching.} 
%

\begin{figure}[t!]
\centering
\includegraphics[width=1.0\linewidth]{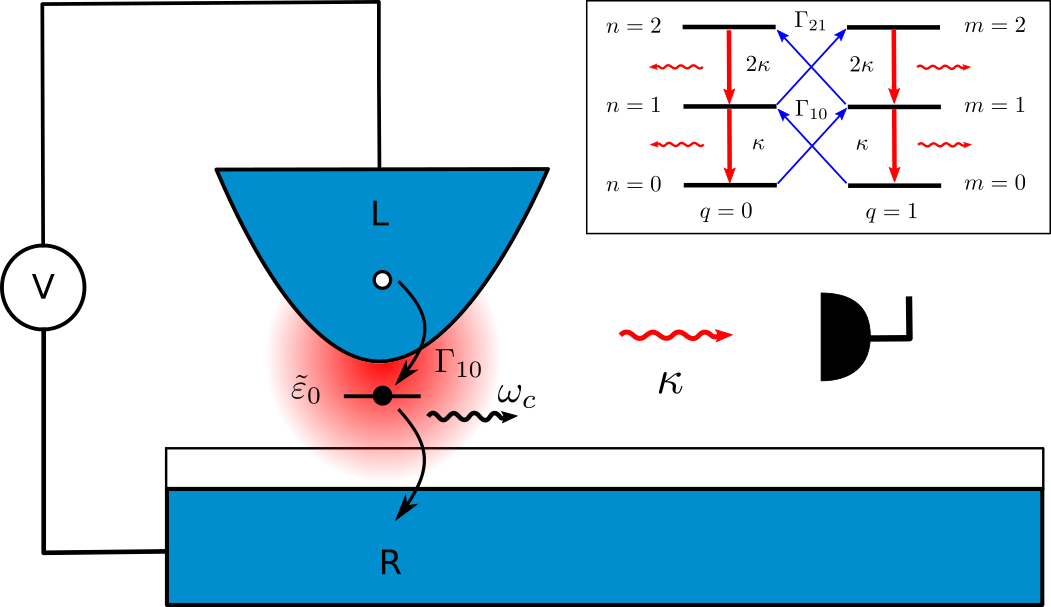}
\caption{
Representation of a current-driven STM plasmonic nanojunction.
The molecule is shown as a single electronic level of energy $\tilde{\varepsilon}_0$.
In presence of a bias-voltage $V$, electrons from the STM apex (L) or from the substrate (R) can tunnel to the electronic level, emitting a cavity plasmon of frequency $\omega_c$.
The former decays with rate $\kappa$, and a photon is emitted (red wavy arrow) that is finally collected by a detector (in black) with detection-yield $\eta$.
Inset: Scheme of the rate equation for the occupation probabilities $P_{(q,n)}(t)$ of the dot charge and cavity plasmonic states $(q,n)$.
The dominant transition rates for the parameters of \Figref{fig:Fig2} are shown.
Thick red (thin blue) lines correspond to the dominant cavity-damping (sub-dominant inelastic single-electron tunneling) processes.
}
\label{fig:Fig1}
\end{figure}
%
%
\textit{Stochastic model of photon-emission}.---We consider the model of 
Ref.~\onlinecite{PhysRevLett.123.246601} describing a single electronic-level molecule embedded inside a STM nanojunction, and coupled to a cavity-plasmon mode (see Fig.\ref{fig:Fig1}).
The state of the plasmon-molecule subsystem is described by two indices $i\equiv (q,n)$ corresponding respectively to the charged (uncharged) dot-level $q=0(1)$, and occupancy state $n \in \mathbb{N}$ of the localized plasmon-mode (see inset of Fig.\ref{fig:Fig1}).
We consider the regime of sequential electronic tunneling and moderate cavity-damping $\Gamma \ll \kappa \leq k_BT/\hbar$ \cite{PhysRevLett.123.246601}, with $T$ the temperature of the leads, $\hbar$ the reduced Planck constant and $k_B$ the Boltzmann constant.
In this regime, the dynamics of the probability $P_i(t)$ of occupying the state $i$ is given by a rate-equation $\dot{P}_i(t) = \sum_j \Gamma_{ij}P_j(t)-\sum_j \Gamma_{ji}P_i(t)$, with $\Gamma_{ij}$ the incoherent rate for the transition $j \rightarrow i$.
We consider two types of rates. 
The first one involves transitions which change the charge-state of the dot $(q,n) \rightarrow (\bar{q}=1-q,m)$ and modify by $m-n$ the occupancy of the cavity-plasmon mode.
We associate to these transitions the corresponding inelastic tunneling rate of single-electrons across the junction: 
$\Gamma_{(\bar{q},m)(q,n)}=\sum_{\alpha} \Gamma_{\alpha} f_{q}\left( \Delta_{mn,\alpha} \right) |\left\langle n | \tilde{m} \right\rangle|^2$, 
with $\Gamma_{\alpha=L}$ the tunneling rate of electrons from the STM apex (L) lead to the dot, and $\Gamma_{\alpha=R}$ the tunneling rate from the substrate (R) lead to the dot.
The factor $|\left\langle n | \tilde{m} \right\rangle|^2$ is the Franck-Condon overlap \cite{PhysRevB.74.205438} between the state $\ket{n}$ of the cavity with empty dot and the displaced-state $\ket{\tilde{m}}$ of the cavity with occupied dot.
We introduced the functions $f_{q=0}\left( E \right) \equiv f(E)$ and 
$f_{q=1}\left( E \right) \equiv 1 - f(E)$, with $f(E)=\left\lbrace e^{E/k_B T} + 1 \right\rbrace^{-1}$ the Fermi-Dirac distribution of the electrons populating the leads.
This function is evaluated at the transition energy $\Delta_{mn,\alpha} = \tilde{\varepsilon}_0 +
 \left( m - n \right)\hbar\omega_c - \mu_{\alpha}$, with $\tilde{\varepsilon}_0 = \varepsilon_0 - \lambda^2 \hbar\omega_c$ the molecular dot-level energy renormalized by \textcolor{black}{its coupling $\lambda$ to the cavity-mode expressed in units of $\hbar\omega_c$ \cite{PhysRevLett.123.246601}}, $\omega_c$ the cavity-plasmon frequency, and $\mu_{\alpha}$ the chemical potential of lead $\alpha$.
The second type of rates involves transitions which do not change the charge state of the dot $(q,n) \rightarrow (q,n-1)$ and decrease by one the number of cavity-plasmons. 
Those incoherent transitions are associated to the cavity-photon losses $\Gamma_{(q,n-1)(q,n)}=\kappa n$, with $\kappa$ the cavity damping-rate at the origin of photon emission by the nanojunction. 
%

%
\begin{figure}[t!]
\centering
\includegraphics[width=1.0\linewidth]{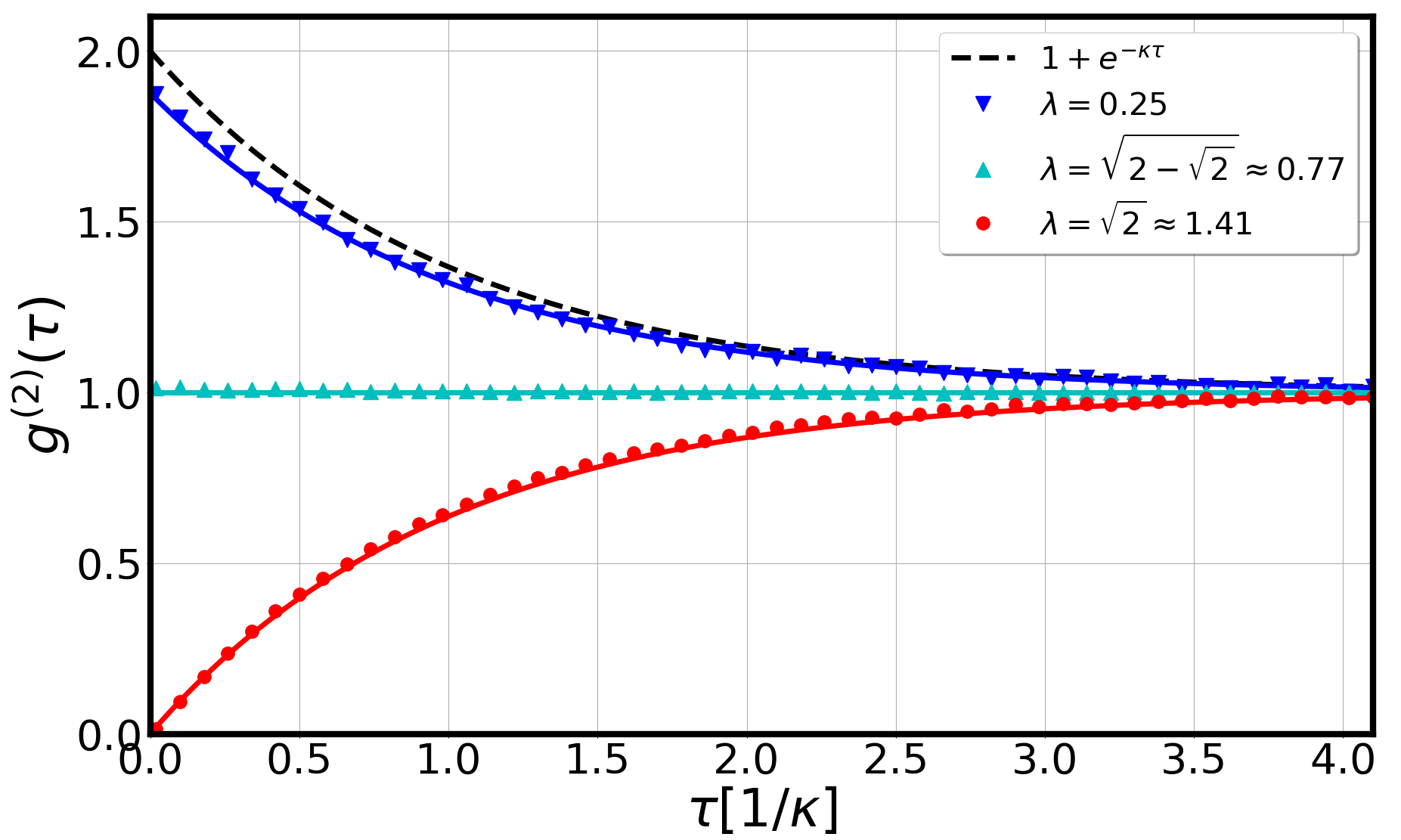}
\caption{
Second-order correlation function $g^{(2)}(\tau)$ for the emitted photons as a function of time $\tau$, obtained numerically from the Monte Carlo simulations (averaged over 40 runs). 
Blue lower triangles are obtained for a plasmon-molecule coupling strength $\lambda=0.25$, cyan upper triangles for $\lambda=\sqrt{2-\sqrt{2}}\approx 0.77$ and red dots for $\lambda=\sqrt{2}\approx 1.41$.
Plain curves are the analytical results from \Eqref{Analyticalg21}.
Parameters are : $\eta=1$, $\kappa=k_B T/\hbar=0.1 \omega_c$, $\Gamma_L=\Gamma_R=\Gamma=0.01 \omega_c$, $\mu_L=-\mu_R=eV/2=\hbar\omega_c$, $\tilde{\varepsilon}_0=0$.
}
\label{fig:Fig2}
\end{figure}
%
%
\textit{Monte Carlo approach}.---We solve numerically the previous rate equation, using a kinetic Monte Carlo (MC) approach \cite{fichthorn1991theoretical,gillespie1976general}.
\textcolor{black}{
We assign a probability, or detection yield, $\eta$ for each photon that has been emitted by the nanojunction to be finally collected and detected by an external photon detector, a perfect detection-yield meaning $\eta=1$.
We suppose in the MC calculation that the photon-detection event by the photon detector is independent from the photon-emission event by the junction \cite{SupMat}.
}
The output of the MC enables to record the history of random times at which a photon is emitted and detected.
From those time-traces we extract $S(\tau) \equiv P\left( \tau| 0\right)$ the conditional probability distribution that a photon is emitted and detected at time $\tau$, knowing that a photon has been emitted and detected at time $0$.
Similarly, we obtain $Q\left( \tau| 0\right)$, the probability distribution of the first-time photon detection event.
It is defined as the exclusive conditional probability distribution of a photon emission and detection event at time $\tau$, knowing that the previous photon detection event occurred at time $0$, \textit{with the constraint that no other photon was emitted in the time interval $\left\rbrack 0, \tau \right\lbrack$.}
The probability distributions $P$ and $Q$ are different, since the first one collects all possible photon emission and detection events in the intermediate time-interval $\left\rbrack 0, \tau \right\lbrack$, while the second one excludes them all. 
\textcolor{black}{The SCF and WTD of emitted and detected photons are obtained 
from those fundamental distributions as \cite{SupMat}}
\begin{eqnarray} 
g^{(2)}(\tau) &=& \frac{S(\tau)}{\Gamma_{\gamma}^{(\rm{st})}}
\, , \label{g2MonteCarlo} \\
w(\tau) &=& Q\left( \tau| 0\right)
\, , \label{WTDMonteCarlo} 
\end{eqnarray} 
with $\Gamma_{\gamma}^{(\rm{st})}=\eta\kappa \left\langle n \right\rangle$ the rate or stationary probability per unit of time to emit and detect a photon, and $\left\langle n \right\rangle$ the average occupation of the cavity-plasmon mode.
%

\begin{figure}[t!]
\centering
\includegraphics[width=1.0\linewidth]{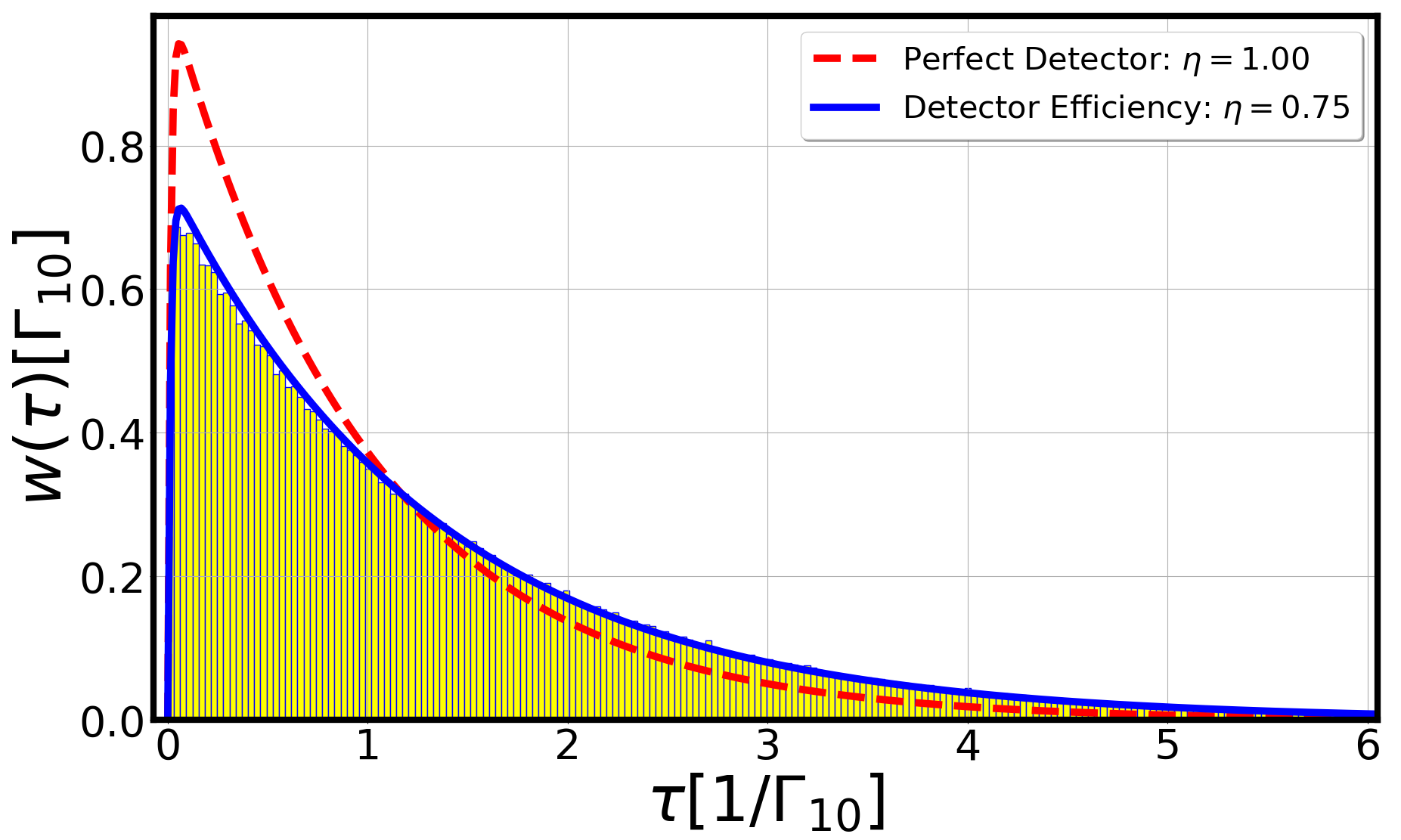}
\caption{
Time-dependence of the distribution of delay-times $w(\tau)$ between two successive photon emission and detection events, obtained from the MC numerical calculation (yellow histogram), expressed in units of $\Gamma_{10}$.
The blue plain (dashed red) curve is the outcome of the analytical formula in \Eqref{AnalyticalWTDAntibunching}, for $\eta=0.75(1)$.
Parameters are those of \Figref{fig:Fig2} with $\lambda=\sqrt{2}$, for which photon antibunching occurs.
}
\label{fig:Fig3}
\end{figure}
%
%
%
\textit{Expressions for the SCF and WTD}.---In the following, we write $P_{mn}^{q'q}\left(\tau\right)$ the occupation probability of the plasmon-molecule state $(q'm)$ at time $\tau$ that is solution of the rate equation, with the state $(q n)$ initially occupied. 
Similarly, we note $Q_{mn}^{q'q}\left(\tau\right)$ the exclusive probability of first reaching the state $(q'm)$, leading to a first-photon emission and detection event at time $\tau$, knowing that the state $(q n)$ was occupied at time $\tau=0$.
From Eqs.~\eref{g2MonteCarlo} and \eref{WTDMonteCarlo}, we derive the following expressions for the SCF and WTD (see Supplementary Material \cite{SupMat} for further details)
\begin{eqnarray} 
S(\tau) &=& \frac{\kappa \eta}{\left\langle n \right\rangle}\sum_{n,m = 1}^{+\infty}\sum_{q,q'=0,1} m n
P_{mn-1}^{q' q}\left(\tau\right)
P^{(\rm{st})}_{(q n)}
\, , \label{g2formal} \\
w(\tau) &=& \frac{\kappa \eta}{\left\langle n \right\rangle}\sum_{n,m = 1}^{+\infty}\sum_{q,q'=0,1} m n
Q_{mn-1}^{q' q}\left(\tau\right)
P^{(\rm{st})}_{(q n)}
\, , \label{WTDformal} 
\end{eqnarray} 
with $P^{(\rm{st})}_{(q n)}=\lim_{\tau\rightarrow +\infty}P_{(q n)}\left(\tau\right)$ the stationary occupancy of the state $(qn)$.
The $Q$-distribution is then solution of a renewal-like integral equation \cite{carmichael1989photoelectron}
\begin{eqnarray} 
P_{nm}^{q q'}\left(\tau\right) &=& 
Q_{nm}^{q q'}\left(\tau\right)
+ \kappa \sum_{k=1}^{+\infty} k \sum_{r=0,1} 
\left( P_{nk-1}^{q r} \ast 
Q_{km}^{r q'} \right)\left(\tau\right)
\, , \nonumber \\
\label{RelationgPQ} 
\end{eqnarray} 
where we wrote $\left( g \ast h \right) \left(\tau\right) \equiv 
\int_0^\tau d\tau_1 g\left(\tau-\tau_1\right) h\left(\tau_1\right)$ the convolution between any two causal functions $g$ and $h$. 
In general, Eq.~\eref{RelationgPQ} has to be solved numerically, after Laplace transforming.
%

\begin{figure}[t!]
\centering
\includegraphics[width=1.0\linewidth]{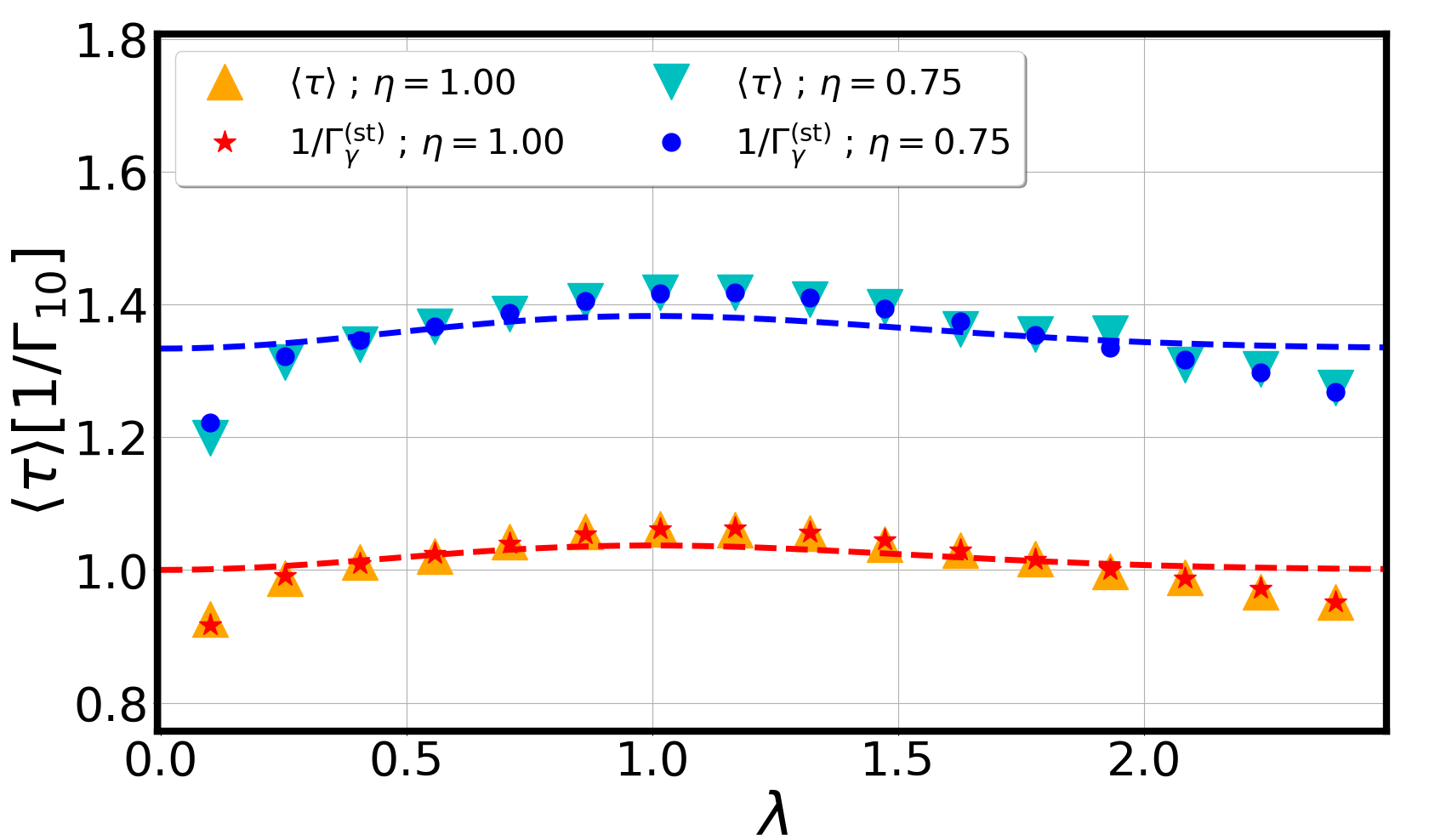}
\caption{
Average-delay time $\left\langle \tau \right\rangle$ in units of $\Gamma_{10}$ as a function of plasmon-molecule coupling $\lambda$, obtained from the MC numerical calculation. 
Upper orange (lower cyan) triangles correspond to the case of a perfect (non-perfect) detection yield $\eta=1(0.75)$.
Red stars (blue dots) are the corresponding values of $1/\Gamma_{\gamma}^{(\rm{st})} \equiv 1/\eta\kappa \left\langle n \right\rangle$ appearing in \Eqref{AverageTimeDelayAntibunching1}.
Dashed curves are the values of $\left\langle \tau \right\rangle$ given by approximate \Eqref{AverageTimeDelayAntibunching2}.
}
\label{fig:Fig4}
\end{figure}
%
%
%
\textit{Results for the SCF}.---In the rest of the paper, we consider the case of an electron-hole symmetric junction for which $\tilde{\epsilon}_0=0$, $\Gamma_L = \Gamma_R = \Gamma$, and $\mu_L=-\mu_R=eV/2$, with $V$ the bias-voltage between source and drain and $e$ the elementary charge. 
In this regime, the inelastic tunneling rates are independent of the charge state, namely $\Gamma_{(\bar{q},m)(q,n)}=\Gamma_{mn}$ for all $q=0,1$.
The general case of asymmetric junctions regarding the SCF is considered in details in Ref.~\onlinecite{PhysRevLett.123.246601}.
In \Figref{fig:Fig2}, we show $g^{(2)}(\tau)$ as a function of time $\tau$, obtained from the MC simulations (averaged over 40 runs). 
The voltage-bias is fixed at the first inelastic threshold for photon-emission ($eV = 2 \hbar \omega_c$) and the photon detection is perfect ($\eta=1$).
Upon increasing the plasmon-molecule coupling strength $\lambda$, we find a crossover in the SCF from photon-bunching to photon-antibunching.
This result agrees with the results found in Ref.~\onlinecite{PhysRevLett.123.246601}, derived with another method.
In this range of parameters, the rate equation for $P_{(q n)}(t)$ is well approximated by truncating the available cavity-occupancies to $n \leq 2$ \cite{PhysRevLett.123.246601}. 
The dominant transition rates are provided by the cavity-damping rate $\kappa$ (red downward arrows in inset of \Figref{fig:Fig1}) and two single-electron inelastic tunneling rates $\Gamma_{10},\Gamma_{21} \ll \kappa$ (blue upward arrows).
This truncated rate equation can be solved analytically exactly \cite{SupMat}, to provide in the regime $\kappa \gg \Gamma$
\begin{eqnarray} 
g^{(2)}(\tau) &\approx & 1 + e^{-\kappa \tau} \left( 
g^{(2)}(0) - 1
\right)
\, , \label{Analyticalg21} \\
g^{(2)}(0) &=& \frac{\langle n(n-1) \rangle}{\langle n \rangle^2}
\approx \frac{\Gamma_{21}}{\Gamma_{10}} = \frac{\left( 2 - \lambda^2 \right)^2}{2}
\, .
\label{Analyticalg22} 
\end{eqnarray} 
The analytical results of \Eqref{Analyticalg21} are shown as plain curves in \Figref{fig:Fig2}, and perfectly agree with the numerically exact MC.
We thus confirm quantitatively the results of Ref.~\onlinecite{PhysRevLett.123.246601} that $g^{(2)}(\tau)$ relaxes exponentially in time towards unity, with a rate given by the cavity-damping rate $\kappa$.
The convergence of $g^{(2)}(\tau)$ to $1$, is due to the fact that two distinct photon emission events separated by a time-interval $\tau \gg 1/\kappa$ become independent.
The zero-delay behavior $g^{(2)}(0)$ between two emission events is given by \Eqref{Analyticalg22}.
For weak plasmon-molecule coupling $\lambda = 0.25 < 1$ (blue lower triangles), the stationary probability of having $n=2$ occupancy of the cavity-plasmon mode is significant.
This results in an effective (out-of-equilibrium) thermal state characterized by photon bunching ($g^{(2)}(\tau) \leq g^{(2)}(0)=2$).
For a critical value of $\lambda = \sqrt{2}$ (red points), the rate $\Gamma_{21}$ vanishes due to the vanishing of the Franck-Condon matrix element. 
This results in a vanishing probability to reach the $n=2$ occupancy, with only two possible occupations of the plasmon-mode $n=0,1$. 
This leads to $g^{(2)}(\tau) \geq g^{(2)}(0)=0$ and thus to photon antibunching, a fingerprint of single-photon emission.
Finally the crossover region that is characterized by $g^{(2)}(\tau) = g^{(2)}(0)=1$ \textcolor{black}{(Poissonian behavior)}, is reached for $\lambda = \sqrt{2-\sqrt{2}}$ (cyan upper triangles).
\textit{Results for the WTD}.---We now consider the time-evolution of $w(\tau)$. 
The average delay-time between two successive photon emission and detection events $\left\langle \tau \right\rangle = \int_0^{+\infty} d\tau \tau w(\tau)$ can be derived analytically from Eqs.~\eref{WTDformal} and \eref{RelationgPQ}, using a similar approach to the one used in computing polymer mean reaction times \cite{guerin2012non,guerin2013reactive,PhysRevE.87.032601,condamin2008probing}.
We obtain the general relation (see \cite{SupMat} for details)
\begin{eqnarray} 
\left\langle \tau \right\rangle &=& \frac{1}{\Gamma_{\gamma}^{(\rm {st})}} \equiv \frac{1}{\eta\kappa \left\langle n \right\rangle}
\, , \label{AverageTimeDelayAntibunching1} 
\end{eqnarray} 
which relates the average cavity-photon occupation $\left\langle n \right\rangle$ to the ratio between the dissipation-time $1/\kappa$ and the average delay-time $\left\langle \tau \right\rangle$.
This relation is reminiscent of Kac's lemma \cite{kac1947notion,aldous2002reversible}.
It is expected to hold in any ergodic system, but as far as we know, Eq.~\eref{AverageTimeDelayAntibunching1} was not clearly identified before in the field of plasmonics. 
From now on, we focus on the case $\lambda=\sqrt{2}$, for which maximal antibunching occurs.
We show in \Figref{fig:Fig3}, the WTD computed numerically with the MC (yellow histogram), in the case of a non-perfect detection yield $\eta=0.75$.
We obtain that $w(\tau)$ is a non-monotonous function of time, with a maximum at times $\tau \approx 1/\kappa$. 
In the same region of parameters for which \Eqref{Analyticalg21} was derived, we obtain \cite{SupMat} 
\begin{eqnarray} 
S(\tau) &=& \frac{\eta\kappa\Gamma_{10}}{\kappa_t} \left\lbrace
1 - e^{-\kappa_t \tau} 
\right\rbrace
\, , \label{AnalyticalSAntibunching} \\
w(\tau) &=& \frac{\eta\kappa\Gamma_{10}}{\kappa_d} \left\lbrace
e^{-\frac{\left(\kappa_t - \kappa_d\right)\tau}{2}} - e^{-\frac{\left(\kappa_t + \kappa_d\right)\tau}{2}}
\right\rbrace
\, , \label{AnalyticalWTDAntibunching} 
\end{eqnarray} 
with $\kappa_d = \sqrt{ \kappa_t^2 - 4\eta\kappa\Gamma_{10}}$, and $\kappa_t = \kappa + \Gamma_{10}$.
Equation \eref{AnalyticalWTDAntibunching} is one of the main results of this paper.
Its outcome is plotted as a plain(dashed) line in \Figref{fig:Fig3} for $\eta = 0.75 (1)$, and matches very well the MC histogram.
At short delay-times ($\tau \ll 1/\kappa$), the antibunching mechanism implies that the probability of emitting two successive photons in a short delay-time $\tau$ is strongly reduced.
The corresponding linear vanishing of $w(\tau) \approx \eta \kappa \Gamma_{10} \tau$ has a slope proportional to $\Gamma_{10}$.
This is due to the fact that after the first emission event, inelastic tunneling of a single-electron across the nanojunction is necessary to emit another cavity-plasmon, that will later decay through a photon-emission event with rate $\kappa$.
The slope also decreases with $\eta$, since it becomes less probable to detect the emitted photon upon worst detection-yield. 
At large delay-times ($\tau \gg 1/\kappa$), the WTD vanishes exponentially as $w(\tau)\approx \eta \Gamma_{10} e^{-\eta \Gamma_{10}\tau}$.
This reflects the fact that a long time $\tau$ after the first emission event, it becomes very unlikely that another photon has not been  emitted.
At intermediate times ($\tau \approx 1/\kappa$), the maximum WTD is reached at a time $\tau_m$ such that
\begin{eqnarray} 
\tau_m &=& \frac{1}{\kappa_d} \ln
\left(\frac{\kappa_t + \kappa_d}{\kappa_t - \kappa_d}\right)
\approx \frac{1}{\kappa}\ln \left(
\frac{\kappa}{\eta \Gamma_{10}}
\right)
\, . \label{AnalyticalWTDMax} 
\end{eqnarray} 
The average delay-time $\left\langle \tau \right\rangle$ results from Eq.~\eref{AnalyticalWTDAntibunching}  
\begin{eqnarray} 
\left\langle \tau \right\rangle &=& \frac{1}{\eta} \left\lbrace 
\frac{1}{\Gamma_{10}} + \frac{1}{\kappa} \right\rbrace
\, , \label{AverageTimeDelayAntibunching2} 
\end{eqnarray} 
and recovers the result of \Eqref{AverageTimeDelayAntibunching1}, in the particular case $\lambda=\sqrt{2}$.
The average delay-time $\left\langle \tau \right\rangle \approx  1/\eta\Gamma_{10}$ is thus proportional to the inelastic tunneling time of single-electrons across the junction $1/\Gamma_{10}$, corresponding to the necessary waiting-time needed for two successive current-driven photon-emission events to occur.  
As expected, the lower the detection-yield, the longer $\left\langle \tau \right\rangle$.
We show in Fig.\ref{fig:Fig4} the robustness of Eq.~\eref{AverageTimeDelayAntibunching1} away from the specific case $\lambda = \sqrt{2}$, for $\eta=0.75$ and $1$.
The quantity $\left\langle \tau \right\rangle$ (lower cyan triangles) obtained from the MC, and $1/\Gamma_{\gamma}^{(\rm{st})}$ (blue dots) derived from solving for the stationary state in the rate equation, are shown to coincide as a function of $\lambda$, for $\eta=0.75$.
The same good agreement is found for $\eta=1$.
Surprisingly, in the range of moderate to strong plasmon-molecule coupling strengths ($\lambda \in \left\lbrack 0.1,2.0 \right\rbrack$), Eq.~\eref{AverageTimeDelayAntibunching2} is still a good approximation to the exact value of $\left\langle \tau \right\rangle$ (see blue dashed curve in Fig.\ref{fig:Fig4}), despite a strong modulation of the rate $\Gamma_{10}$ with $\lambda$.
Furthermore, we note that an integral equation exists relating $S(\tau)$ and $w(\tau)$ \cite{SupMat},
\begin{eqnarray} 
S(\tau) &=& w(\tau) + \left( S \ast w \right) (\tau)
\, , \label{Relationg2AndWTD} 
\end{eqnarray} 
that is consistent with Eqs.~\eref{AnalyticalSAntibunching}-\eref{AnalyticalWTDAntibunching}.
Equation \eref{Relationg2AndWTD} was derived previously in Refs.~\onlinecite{reynaud1983fluorescence,PhysRevLett.69.1516} for describing the stochastic Markovian dynamics of fluorescent two-level atoms or molecules.
In our case, however, this relation is valid only at electron-hole symmetric point for $eV=2\omega_c$ and $\lambda=\sqrt{2}$, for which only two cavity states $n=0,1$ matter.
In general, for arbitrary values of external parameters, the validity of this relation is not granted anymore, and one resorts with either MC simulations, or with solving numerically the linear system of Eqs.~\eref{RelationgPQ} to obtain the $Q$-distribution and the WTD in Eq.~\eref{WTDformal}.
\textcolor{black}{Finally, we remark that Eqs.~\eref{Analyticalg21} and \eref{AverageTimeDelayAntibunching2} resolve the timescale issue noticed in the introduction.}
In the regime $\Gamma \ll \kappa$, there is no contradiction having $g^{(2)}(\tau)$ relaxing exponentially with the cavity-damping time $1/\kappa$, while the average delay-time $\left\langle \tau \right\rangle$ is given by the inverse inelastic tunneling time $1/\Gamma_{10}$ of single-electrons across the nanojunction. 
This difference of timescales is due to the fact that the SCF and WTD  do not provide the same information about the statistics of photon emission and detection, and should thus be seen as complementary statistical indicators.
%

%
\textit{Conclusions}.---We have investigated in depth the time-dependence of the second-order correlation function $g^{(2)}(\tau)$ and waiting-time distribution $w(\tau)$ of photons emitted by a current-induced plasmonic nanojunction with a  single electronic level.
By using MC and analytical calculations, 
we have shown that the two quantities provide a complementary information about the statistics of emitted photons by the nanojunction.
In the regime of photon-antibunching, and when $\kappa \gg \Gamma$, $g^{(2)}(\tau)$ relaxes in time towards unity with the cavity damping-time $1/\kappa$, while the average delay-time $\left\langle \tau \right\rangle$ between successive photon emission and detection events is proportional to the inelastic tunneling time of single-electrons across the nanojunction $1/\Gamma_{10}$.
We hope that our paper will stimulate further experiments in current-driven STM plasmonic nanojunctions, for which, to the best of our knowledge, a comparative measurement of the WTD with respect to the SCF of the emitted photons is still lacking, but seems to be crucial to understand better the timescales and physical mechanism involved in the current-induced light-emission process.
%

We acknowledge fruitful discussions with Guillaume Schull, Benoit Douçot and Javier Aizpurua about the statistics of photon emission, and with Ludovic Jaubert about kinetic MC simulations. We also thank Thomas Gu\'{e}rin for pointing out Kac's lemma in our interpretation of the average delay-time between two photon emission events. 
This work was supported by IDEX Bordeaux (No.~ANR-10-IDEX-03-02) and Euskampus Transnational Common Laboratory \textit{QuantumChemPhys}.
R.~A.~acknowledges financial support by the French Agence Nationale de la Recherche project CERCa, ANR-18-CE30-0006.
T.~F.~acknowledges financial support by the Spanish AEI (FIS2017-83780-P \textcolor{black}{and PID2020-115406GB-I00}) and the European Union’s Horizon 2020 (FET-Open project SPRING Grant No.~863098).
F.~P. acknowledges support from the French Agence Nationale de la Recherche (grant SINPHOCOM ANR-19-CE47-0012).
\bibliography{Biblio_Monte_Carlo}
\end{document}


\title{Supplementary Material to the Paper: \mbox{"Photon-Emission Statistics induced by Electron Tunnelling in Plasmonic Nanojunctions"}}

%
\author{R. Avriller}
\affiliation{Univ. Bordeaux, CNRS, LOMA, UMR 5798, F-33405 Talence, France}
\author{Q. Schaeverbeke}
\affiliation{Univ. Bordeaux, CNRS, LOMA, UMR 5798, F-33405 Talence, France}
\affiliation{Donostia International Physics Center (DIPC), E-20018 Donostia-San Sebasti\'{a}n, Spain}
\author{T. Frederiksen}
\affiliation{Donostia International Physics Center (DIPC), E-20018 Donostia-San Sebasti\'{a}n, Spain}
\affiliation{Ikerbasque, Basque Foundation for Science, E-48013, Bilbao, Spain}
\author{F. Pistolesi}
\affiliation{Univ. Bordeaux, CNRS, LOMA, UMR 5798, F-33405 Talence, France}
\date{\today}

\begin{abstract}
We provide additional information about the analytical calculations performed in the paper, giving the second-order correlation functions of emitted photons $g^{(2)}(\tau)$, and the delay-time distribution between two photon emission events $w(\tau)$. 
We derive the general expression for the average delay-time $\left\langle \tau \right\rangle$ between two photon-emission events from the rate equation.
\end{abstract}

\maketitle

\tableofcontents

\section{Considerations about the rate equation}
\label{RE_Considerations}
\subsection{Rate equation for the symmetric junction}
\label{RE_Considerations1}
We describe the dynamics of states $(q,n)$, with $q=0(1)$ the charge-state of the empty(filled) dot and $n \in \mathbb{N}$ the number of photons inside the STM plasmonic cavity.
The time-dependent probability of occupying those states $P_{(q,n)}(t)$ is solution of the rate-equation (RE)
\begin{eqnarray}
\dot{P}_{(q,n)}(t) &=& \sum_{m=0}^{+\infty} \left\lbrace
\Gamma_{n m}^{q \overline{q})}P_{(\overline{q},m)}(t)
-
\Gamma_{mn}^{\overline{q} q} P_{(q,n)}(t)
\right\rbrace \nonumber \\
&+& (n+1)\kappa_\downarrow P_{(q,n+1)}(t) + n\kappa_\uparrow P_{(q,n-1)}(t) \nonumber \\ 
&-& \left\lbrack (n+1)\kappa_\uparrow + n\kappa_\downarrow \right\rbrack
P_{(q,n)}(t) \, ,
\label{RE_1} 
\end{eqnarray}
with given initial conditions
$P_{(q,n)}(0)$.
\Eqref{RE_1} is simplified in the case of an electron-hole symmetric junction, for which $\tilde{\varepsilon}_0=\varepsilon_0-\lambda^2\hbar\omega_c=0$, $\Gamma_L=\Gamma_R=\Gamma$, and $\mu_L=-\mu_R=eV/2$.
Indeed in this regime, the tunneling rates are independent of the charge-states of the dot
\begin{eqnarray}
\Gamma_{m n}^{01} &=&  \Gamma_{m n}^{10} \equiv \Gamma_{mn} 
\label{RE_2} \, , \\
\Gamma_{mn} &=& \Gamma |\langle m | \tilde{n}  \rangle |^2 \sum_{\alpha=\pm} f\left\lbrack \left( m - n \right)\hbar\omega_c - \alpha \frac{eV}{2} \right\rbrack \, .
\end{eqnarray}
Moreover, the fact that $T \ll \hbar\omega_c/k_B$, implies that $\kappa_\uparrow \approx 0$ and $\kappa_\downarrow \approx \kappa$.
Using the former simplification and \Eqref{RE_2}, the occupation probability of the cavity-mode $\pi_n(t)=\sum_{q=0,1}P_{(q,n)}(t)$ which is obtained after integrating-out the dot-charge states, is shown to be solution of a simplified RE
\begin{eqnarray}
\dot{\pi}_{n}(t) &\approx & \sum_{m=0}^{+\infty} \left\lbrace 
\Gamma_{n m} \pi_{m}(t) - \Gamma_{m n}\pi_{n}(t)
\right\rbrace
\nonumber \\
&+& \kappa (n+1)\pi_{n+1}(t) - \kappa n \pi_{n}(t)
\label{RE_3} \, ,
\end{eqnarray}
with initial conditions $\pi_{n}(0)$.
\subsection{Decomposition of the RE into eigenmodes}
\label{RE_Considerations2}
We define $\underline{\pi}(t)$ the vector of which components are given by $\left\lbrack\underline{\pi}(t)\right\rbrack_{n}\equiv \pi_n(t)$. 
We note $\mathbf{\Gamma}$ the rate-matrix such that \Eqref{RE_3} can be written in vectorial form $\dot{\underline{\pi}}(t)=\mathbf{\Gamma}\underline{\pi}(t)$, with initial condition $\underline{\pi}(0)$.
The solution of the RE is then given by $\underline{\pi}(t)=e^{\mathbf{\Gamma}t}\underline{\pi}(0)$.
We can expand this solution into eigenmodes of the RE. 
For this purpose, we define the right and left eigenvectors $\underline{v}_\lambda$ and $\underline{w}_\lambda$ associated to the $\lambda$ (real and negative) eigenvalue of the $\mathbf{\Gamma}$-matrix, such that $\mathbf{\Gamma}\underline{v}_\lambda = \lambda \underline{v}_\lambda$ and 
$\underline{w}^t_\lambda\mathbf{\Gamma} = \lambda \underline{w}^t_\lambda$.
The null eigenvalue $\lambda=0$ is a trivial eigenvalue of the $\mathbf{\Gamma}$-matrix.
The corresponding null right eigenvector $\underline{v}_0=\underline{\pi}^{\rm{(st)}}$ provides the vector of stationary cavity-mode populations $\underline{\pi}^{\rm{(st)}}=\lim_{t\rightarrow+\infty}\underline{\pi}(t)$, solution of $\mathbf{\Gamma}\underline{\pi}^{\rm{(st)}}=\underline{0}$.
The null left eigenvector is the vector $\underline{w}$ with components
$\left\lbrack\underline{w}\right\rbrack_{n}\equiv 1$ associated to the condition of normalisation of the probability distribution, namely that
$\underline{w}^{t}\cdot \underline{\pi}(t) = \sum_{n=0}^{+\infty}\pi_n(t)=1$ at all times $t$.
We get the solution of the RE as a linear superposition of eigenmodes with coefficients given by the initial condition
\begin{eqnarray}
\underline{\pi}(t) &=& \sum_{\lambda}
\left\lbrack \underline{w}^t_\lambda \cdot \underline{\pi}(0)\right\rbrack \underline{v}_\lambda e^{\lambda t}
\label{EigenmodeDecomposition1} \, .
\end{eqnarray}
A specific choice of initial condition $\pi_n(0)=\delta_{nm}$ enables to derive from \Eqref{EigenmodeDecomposition1} the conditional probability $P(nt|m0)$ that the state $n$ of the cavity-mode is realized at time $t$, knowing that the cavity was in state $m$ at initial time $t=0$
\begin{eqnarray}
P(nt|m0) &=& \pi_n^{\rm{(st)}} + \mathcal{G}_{nm}(t)
\label{EigenmodeDecomposition2} \, , \\
\mathcal{G}_{nm}(t) &=& \sum_{\lambda\neq 0} \left\lbrack \underline{v}_{\lambda} \right\rbrack_{n}\left\lbrack \underline{w}_{\lambda} \right\rbrack_{m} e^{\lambda t}
\label{EigenmodeDecomposition3} \, ,
\end{eqnarray}
with $\left\lbrack \mathcal{G}(t) \right\rbrack_{nm}\equiv \mathcal{G}_{nm}(t)$ the pseudo Green-function of the RE.
\Eqref{EigenmodeDecomposition2} is the starting point of the analytical derivation for the average-delay time between two photon emission events (see \Secref{SecWTD2}).
We further define the Laplace-transform of any causal function $g(t)$ written $\tilde{g}(z)=\int_0^{+\infty}dt f(t)e^{-zt}$, and obtain
\begin{eqnarray}
\tilde{P}(nz|m0) &=& \frac{\pi_n^{\rm{(st)}}}{z} + \tilde{\mathcal{G}}_{nm}(z)
\label{EigenmodeDecomposition4} \, , \\
\tilde{\mathcal{G}}_{nm}(z) &=& \sum_{\lambda\neq 0} \frac{\left\lbrack \underline{v}_{\lambda} \right\rbrack_{n}\left\lbrack \underline{w}_{\lambda} \right\rbrack_{m}}{z+\lambda} 
\label{EigenmodeDecomposition5} \, .
\end{eqnarray}
%

\section{Second-order correlation function of the emitted photons}
\label{SCF_General}
\subsection{Monte Carlo approach}
\label{SCF_General1}
\subsubsection{MC algorithm}
\label{SCF_Algorithm}
%
\begin{figure}[t!]
\centering
\includegraphics[width=0.7\linewidth]{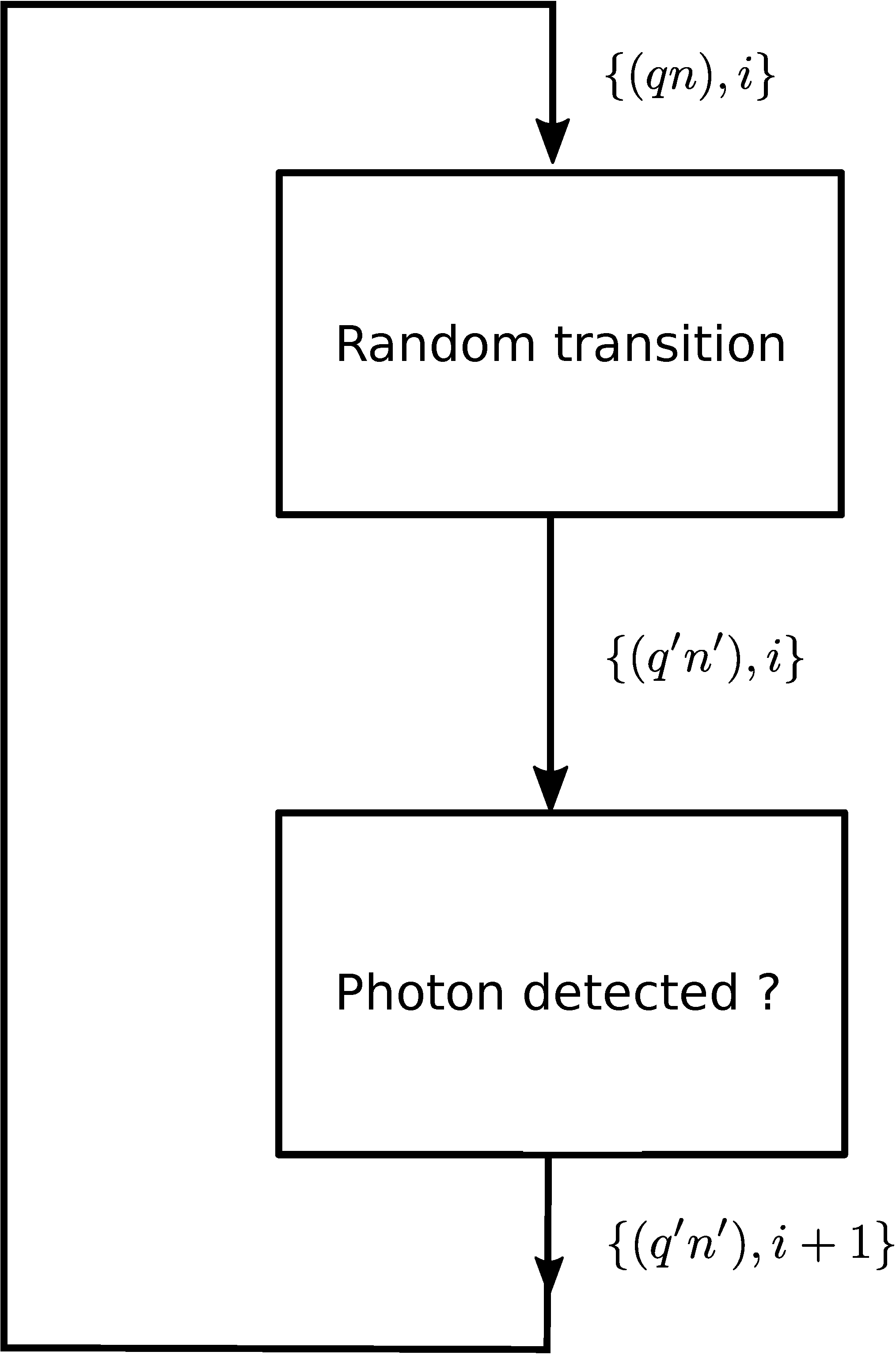}
\caption{
Graphical representation of the kinetic Monte Carlo algorithm, solving numerically the RE of Eq.\refe{RE_1}.
The notation $\left\lbrace (qn), i \right\rbrace$ stands for the list of system states $(qn)$ and times $i$ (in units of $dt$) in the MC loop.
}
\label{fig:Fig1bis}
\end{figure}
%
\begin{figure}[t!]
\centering
\includegraphics[width=1.0\linewidth]{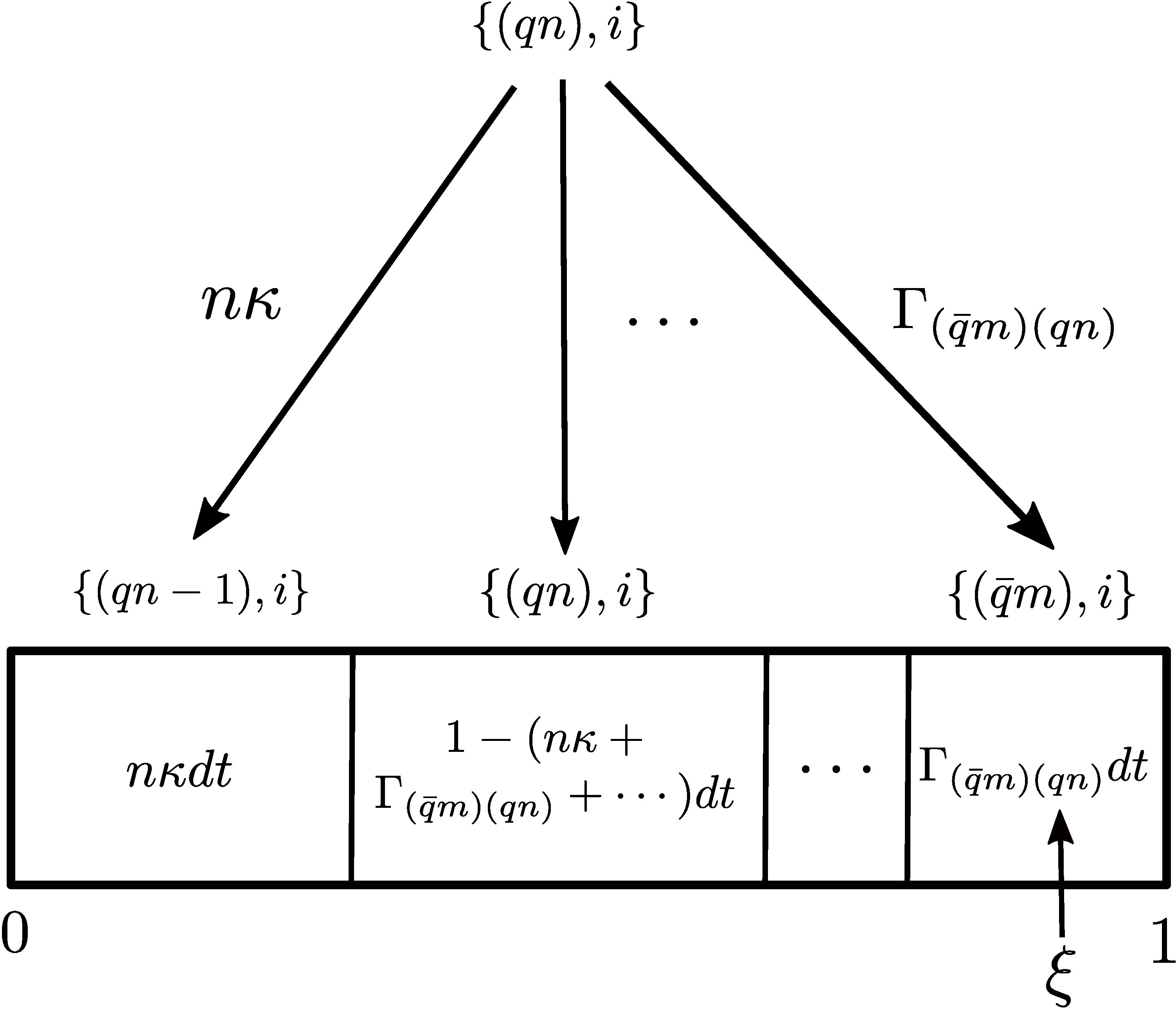}
\caption{
Upper part: Details of the possible random transitions starting from the initial $(qn)$ state in the box \textit{Random transition} of Fig.\ref{fig:Fig1bis}.
Lower part: The random selection of the final state $(q'n')$ is made by separating the interval $\left\lbrack 0, 1 \right\rbrack$ into boxes of lengths proportional to the rate of each transitions starting from $(qn)$.
A random number $\xi \in \left\lbrack 0, 1 \right\rbrack$ is generated that will fall in one box corresponding to the selected state: here $(q'n')=(\bar{q}m)$.
}
\label{fig:Fig1terce}
\end{figure}
%
%
\textcolor{black}{We implement in Python a kinetic Monte Carlo approach \cite{fichthorn1991theoretical,gillespie1976general} to solve numerically for the RE of \Eqref{RE_1}.
A simplified sketch of the MC algorithm is provided in Fig.\ref{fig:Fig1bis}. 
We first discretize the time in $N_t$ equal time steps of size $dt$: typically, in our simulations, $dt\approx 0.02/\kappa$.
The algorithm starts with the list $\left\lbrace (qn), 0 \right\rbrace$ containing an arbitrary initial state of the system $(qn)$ at initial time $0$.
}

\textcolor{black}{
%
%
%
%
%
%
%
We then choose randomly (with the procedure explained in the following) a state $(q'n')$ that is connected to the state $(qn)$ by the RE (see the box labelled \textit{Random transition} in Fig.\ref{fig:Fig1bis}).
Fig.\ref{fig:Fig1terce} (upper part) shows the possible transitions that might happen. 
The transition $(q,n) \rightarrow (q,n-1)$ corresponds to a random event for which "the charge state of the dot is unchanged" and "a photon leaves the cavity" with rate $n\kappa$.
%
%
The transition $(q,n) \rightarrow (\overline{q},m)$ correspond to another random event for which photons are emitted inside the cavity by any inelastic electronic tunneling event, with a rate $\Gamma_{(\bar{q}m)(qn)}$. 
%
%
All the other possible transitions are taken into account, as well as the particular event corresponding to leaving the state of the system unchanged $(q,n) \rightarrow (q,n)$.
In Fig.\ref{fig:Fig1terce} (lower part), we separate the interval $\left\lbrack 0, 1 \right\rbrack$ into boxes of lengths proportional to each of the transitions that might be chosen, for instance one box with length $\propto n\kappa dt$ for the transition $(q,n) \rightarrow (q,n-1)$, another box of length $\propto \Gamma_{(\bar{q}m)(qn)}dt$ for the transition $(q,n) \rightarrow (\overline{q},m)$.
The event corresponding to stay on the same state $(q,n) \rightarrow (q,n)$, has a box of length $\propto 1 - \left( n\kappa + \Gamma_{(\bar{q}m)(qn)} + \cdots \right)dt$.
We finally generate a random number $\xi\in\left\lbrack 0, 1 \right\rbrack$ that will fall in one box of the interval corresponding to the state reached by the system.
For instance, the final state $(q'n')=(\bar{q}m)$ is reached in Fig.\ref{fig:Fig1terce} (lower part).
In our MC simulations, we discard the first $10^6$ time-steps in order to reach the steady-state before recording the time-trace used for the subsequent statistical analysis.
}

\subsubsection{Photon detection}
\label{SCF_Photon_Detection}
\textcolor{black}{
Once a transition to the state $(q'n')$ has occurred, the algorithm selects further if "yes" or "no" this transition is associated to the emission of a photon out of the cavity (see the box labelled \textit{Photon detected ?} in Fig.\ref{fig:Fig1bis}).
Note that, similarly to a photocounting experiment, we consider that only photons emitted through the transitions $(q,n) \rightarrow (q,n-1)$ are recorded in the histories of photon-emission events, namely we record only events at which "the charge state of the dot is unchanged" and "a photon leaves the cavity" with rate $\kappa$ (the emitted photon needs to leave the cavity before being detected).
The photons emitted inside the cavity by any inelastic electronic tunneling event $(q,n) \rightarrow (\overline{q},m)$ (with rate $\Gamma_{mn}$) are thus not recorded. 
In this section, each time we write that a photon is "emitted", we mean that a photon is "emitted out of the cavity".
%
%
%
We then repeat the MC loop of Fig.\ref{fig:Fig1bis} many times ($N_t$ times), so that we reach times scales sufficiently long for improving the precision of the MC calculation and decrease the error bars.
At the output of this calculation, we obtain the history of random times at which a single-photon was emitted out of the cavity.}
\subsubsection{SCF calculation}
\label{SCF_g2}
%
%
%
\begin{figure}[t!]
\centering
\includegraphics[width=1.0\linewidth]{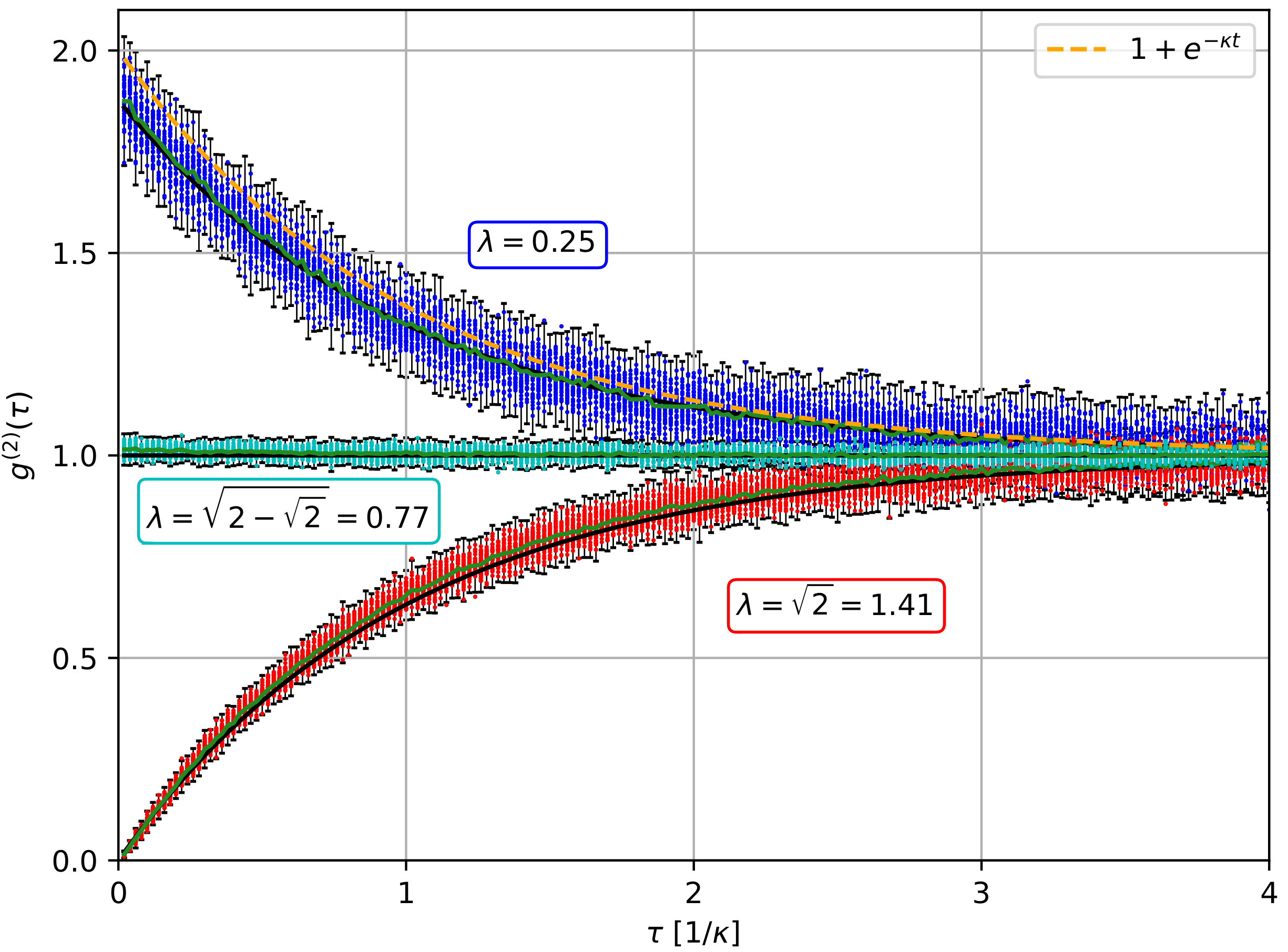}
\caption{
Second-order correlation function $g^{(2)}(\tau)$ computed from the MC calculation. 
A number of $40$ different MC runs are shown as dots to illustrate the uncertainty bar of the method. 
Plain curves are the averages $g^{(2)}(\tau)$ taken over those $40$ MC runs. 
The error bars represent three standard deviations.
Parameters are the same as in Fig.2 of the paper. 
}
\label{fig:Fig1quarte}
\end{figure}
%
%
%
We define the following correlation functions of the emitted photon field
\begin{enumerate}
\item
$G_1(t_1)dt_1 \equiv$ the probability to "emit a photon" in the time interval $\lbrack t_1, t_1 + dt_1 \rbrack$,
\item
$G_2(t_1,t_2)dt_1 dt_2 \equiv$ the joint probability to "emit a first photon" in the time interval $\lbrack t_1, t_1 + dt_1 \rbrack$ and to "emit a second photon" in the time interval $\lbrack t_2, t_2 + dt_2 \rbrack$.
\end{enumerate}
%
%
The normalized second-order correlation function (SCF) of the emitted photons is defined by
\begin{eqnarray}
g^{(2)}(t_1,t_2) = \frac{G_2(t_1,t_2)}{G_1(t_1)G_1(t_2)}
\label{g2_1} \, .
\end{eqnarray}
In the stationary regime, we get that $G_1(t_1) \equiv G_1^{\mathrm{st}}$, $G_2(t_1,t_2) \equiv G_2^{\mathrm{st}}(\tau = t_2-t_1)$ and $g^{(2)}(t_1,t_2)\equiv g^{(2)}(\tau=t_2-t_1)$. 
The stationary SCF is obtained as
\begin{eqnarray}
g^{(2)}(\tau) = \frac{G_2^{\mathrm{st}}(\tau)}{(G_1^{\mathrm{st}})^2}
\label{g2_2} \, .
\end{eqnarray}
\textcolor{black}{
Both quantities $G_1^{\mathrm{st}}$ and $G_2^{\mathrm{st}}(\tau)$ are extracted from the Monte Carlo time-traces in the following way:
i) $G_1^{\mathrm{st}}$ is obtained by the ratio between the total number of photon detection events and the number $N_t$ of time steps,
ii) $G_2^{\mathrm{st}}(\tau)$ is computed as the ratio between the number of pairs of photon emission events delayed by the (discrete) time interval $\tau$ and $N_t$.
Numerically, this is efficiently computed by dividing the whole time-trace into smaller sequences, for which the time differences in each sequence (as well as between pairs of sequences) are sequentially added to a sparse data array that records the number of occurrences for each possible time delay.}
\textcolor{black}{
We present in Fig.\ref{fig:Fig1quarte}, the output of the SCF calculation for $g^{(2)}(\tau)$ for $40$ different MC runs (dot points),  from which error bars of three standard deviations ($99.7\%$ of the values from a normal distribution) are constructed.
This enable to get the typical error bars of the MC calculation.
The average curves over the $40$ MC runs (full lines) are the one used for Fig.2 of the paper.
}

\subsection{Explicit expression of the SCF from the RE}
\label{SCF_General2}
%
\begin{figure}[t!]
\centering
\includegraphics[width=1.0\linewidth]{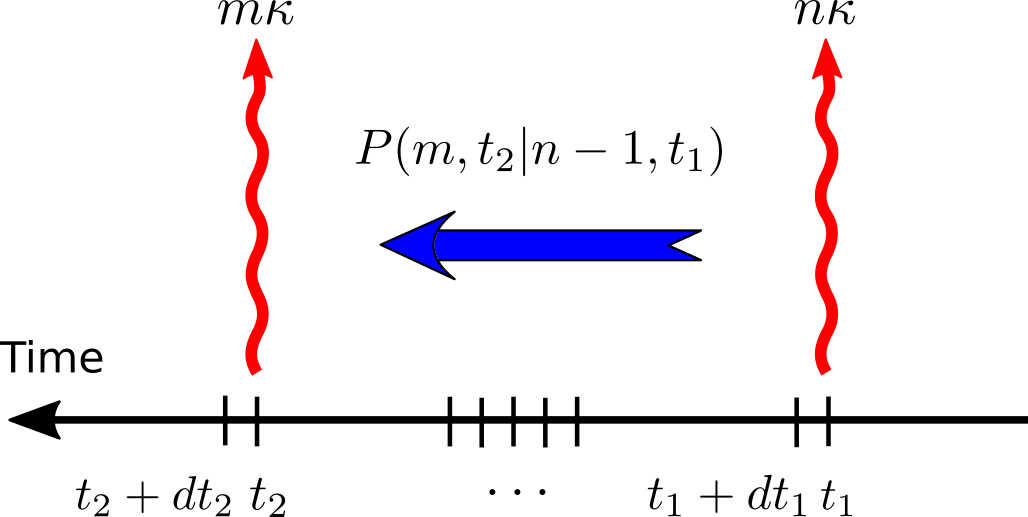}
\caption{
Graphical representation of \Eqref{Glauber_3} giving the unnormalized second-order correlation function for the emitted photons $G^{(2)}(t_1,t_2)$ as a function of time.
The photons emitted at time $t_1$ and $t_2$ are presented as red arrows, while the blue arrow stands for the time-evolution provided by \Eqref{EigenmodeDecomposition2} in between those two detection events.
}
\label{fig:Fig1}
\end{figure}
%
%
In this section, for notational simplicity, we will restrict to the electron-hole symmetric molecular junction, the RE which is given by \Eqref{RE_3}.
The first-order correlation function $G_1(t_1)$ is formally given by 
\begin{eqnarray}
G_1(t_1) &=& 
\sum_{n=0}^{+\infty} n\kappa \pi_{n}(t_1)
\label{Glauber_1} \, .
\end{eqnarray}
Once the stationary state is reached, \Eqref{Glauber_1} becomes
\begin{eqnarray}
G_1^{\rm{st}} &=& \kappa \langle n \rangle \equiv \Gamma_\gamma^{(\rm{st})}
\label{Glauber_2} \, ,
\end{eqnarray}
with $\Gamma_\gamma^{(\rm{st})}$ the rate or stationary probability per unit of time to emit a photon, and $\langle n \rangle = \sum_{n=0}^{+\infty} n \pi_{n}^{\rm{st}}$ the average number of photons inside cavity.
The unnormalized SCF $G_2(t_1,t_2)$ is obtained as (see \Figref{fig:Fig1})
\begin{eqnarray}
G_2(t_1,t_2) 
&=& \sum_{n,m=0}^{+\infty} m\kappa   
P\left( m,t_2|n-1,t_1\right)
n\kappa \pi_{n}(t_1)
\, , \nonumber \\
\label{Glauber_3} 
\end{eqnarray}
from which we obtain, in the stationary state
\begin{eqnarray}
G^{\rm{st}}_2(\tau) &=& \kappa^2
\sum_{n,m=0}^{+\infty} n m 
P\left( m,\tau | n-1,0\right) \pi^{\rm{st}}_{n}
\label{Glauber_4} \, , \\
g^{(2)}(\tau) &=& \frac{1}{\langle n \rangle^2}
\sum_{n,m=0}^{+\infty} n m 
P\left( m,\tau | n-1,0\right) \pi^{\rm{st}}_{n}
\label{Glauber_5} \, .
\end{eqnarray}
\Eqref{Glauber_5} is the main result of this section.
It connects $g^{(2)}(\tau)$ to the conditional probability 
$P\left( m,\tau | n-1,0\right)$ obtained in \Eqref{EigenmodeDecomposition2} from the solution of the RE.
Note that that the state $n-1$ is reached and not the state $n$ in $P\left( m,\tau | n-1,0\right)$.
This is due to the fact that for a single photon to be emitted out of the cavity, a cavity-photon has to be destroyed before.  
Formally, \Eqref{Glauber_5} is equivalent to the output of the numerical Monte Carlo calculation.
\subsection{Relation of the SCF to Glauber formula}
\label{SCF_General3}
In this section, we rewrite the correlation functions in \Eqref{Glauber_2} and \Eqref{Glauber_4} in terms of quantum operators.
We suppose that the stationary reduced density-matrix $\rho^{\rm{st}}$ for the cavity-mode is diagonal in the basis of the $n$ states, namely that 
$\rho^{\rm{st}}=\sum_{n=0}^{+\infty} \pi^{\rm{st}}_{n} \ket{n}\bra{n}$.
Substituting the photon number $n$ by the quantum operator for the photon number $\hat{n} = a^\dagger a$ in \Eqref{Glauber_2}, with $a$ the destruction operator of the cavity-photon mode, we obtain
\begin{eqnarray}
G_1^{\rm{st}} &\equiv & \kappa \mbox{ Tr }  \left\lbrace
a^\dagger a \rho^{\rm{st}}
\right\rbrace
\label{EqGlauber_1} \, .
\end{eqnarray}
In the same way, using the basic postulates of quantum mechanics, we substitute in \Eqref{Glauber_4} the conditional probability $P\left( m,\tau|n-1,0\right)$ by its expression in terms of the cavity-mode wave functions $P\left( m,\tau | n-1,0\right) \equiv  
|\bra{m} e^{-i\hat{H}\tau/\hbar} \ket{n-1}|^2$, with $\hat{H}$ the full Hamiltonian of the STM nanoplasmonic junction.
We obtain
\begin{eqnarray}
G^{\rm{st}}_2(\tau) &=& 
\kappa^2
\sum_{n=0}^{+\infty} 
n \bra{n-1} e^{i\hat{H}\tau/\hbar}
a^\dagger a e^{-i\hat{H}\tau/\hbar}
\ket{n-1} \pi^{\rm{st}}_{n} 
\nonumber \\
&=& \kappa^2
\sum_{n=0}^{+\infty} 
\bra{n} a^\dagger e^{i\hat{H}\tau/\hbar}
a^\dagger a e^{-i\hat{H}\tau/\hbar} a
\ket{n} \pi^{\rm{st}}_{n} 
\nonumber \\
&=& \kappa^2
\mbox{ Tr } 
\left\lbrace
a^\dagger e^{i\hat{H}\tau/\hbar}
a^\dagger a e^{-i\hat{H}\tau/\hbar} a \rho^{\rm{st}}
\right\rbrace 
\label{EqGlauber_2} \, ,
\end{eqnarray}
where we used the relations $a\ket{n}=\sqrt{n-1}\ket{n}$ and $\sum_{m=0}^{+\infty}\ket{m}\bra{m}=\mathbb{Id}$, with $\mathbb{Id}$ the identity matrix.
From \Eqref{EqGlauber_1} and \eref{EqGlauber_2} follows the expression
\begin{eqnarray}
g^{(2)}(\tau) &=& \frac{\mbox{ Tr } 
\left\lbrace
a^\dagger e^{i\hat{H}\tau/\hbar}
a^\dagger a e^{-i\hat{H}\tau/\hbar} a \rho^{\rm{st}}
\right\rbrace}{\mbox{ Tr }  \left\lbrace
a^\dagger a \rho^{\rm{st}}
\right\rbrace^2}
\label{EqGlauber_3} \, .
\end{eqnarray}
\Eqref{EqGlauber_3} recovers the result of Glauber \cite{PhysRev.130.2529} and the expression used in Ref.~\onlinecite{PhysRevLett.123.246601}, for which 
$g^{(2)}(\tau)= \left\langle a^\dagger(0)a^\dagger(\tau)a(\tau) a(0) \right\rangle/\left\langle a^\dagger a \right\rangle^2$.
%

%
\section{Waiting-time distribution of the emitted photons}
\label{SecWTD}
\subsection{Definition and computation of the WTD}
\label{SecWTD1}
In this section, we consider another indicator for the statistics of emitted light out of the cavity, which is the delay-time distribution or waiting-time distribution (WTD) $w(\tau)$ between two successive emitted photons.
For computing the WTD, we first need to introduce the following  distribution functions
\begin{enumerate}
\item
$P(t_2 |t_1)\equiv$ conditional probability distribution to emit a photon at time $t_2$, knowing that a photon has been previously emitted at time $t_1$.
\item
$Q(t_2 | t_1)\equiv$ exclusive conditional probability distribution to emit a photon at time $t_2$, knowing that a photon has been previously emitted at time $t_1$, with the constraint that no photon has been emitted in the time-interval $\rbrack t_1, t_2 \lbrack$.
\end{enumerate}
Both $P$ and $Q$ are computed from the output of the Monte Carlo calculation.
We remark that in general $P\neq Q$.
Indeed, $Q$ involves to the statistics of "exclusive" histories in which no photon-emission event is observed in between two successive photon-emission events, while 
$P$ includes the statistics of the "complete" history of events ("$0$-photon" emission, plus "$1$-photon" emission event, plus "$2$-photon" emission events, etc $\cdots$) in between two photon-emission events.
The SCF and WTD can be expressed in terms of these distributions 
\begin{eqnarray}
g^{(2)}(\tau) &=& \frac{S(\tau)}{\Gamma_\gamma^{(\rm{st})}}
\label{WTD1} \, , \\ 
w(\tau) &=& Q(\tau | 0)
\label{WTD2} \, ,
\end{eqnarray}
with $S(\tau) \equiv P(\tau | 0)$.
\Eqref{WTD1} coincides with \Eqref{g2_2}, while using the same derivation as in \Secref{SCF_General2}, we obtain the following expressions for both $S(\tau)$ and $w(\tau)$
\begin{eqnarray}
S(\tau) &=& \frac{\kappa}{\langle n \rangle}
\sum_{n,m=0}^{+\infty} n m 
P\left( m,\tau | n-1,0\right) \pi^{\rm{st}}_{n}
\label{WTD3} \, , \\
w(\tau) &=& \frac{\kappa}{\langle n \rangle}
\sum_{n,m=0}^{+\infty} n m 
Q\left( m,\tau | n-1,0\right) \pi^{\rm{st}}_{n}
\label{WTD4} \, .
\end{eqnarray}
Introducing the notation $P_{nm}\left(\tau\right)\equiv P\left( n,\tau | n,0\right)$ and $Q_{nm}\left(\tau\right)\equiv Q\left( n,\tau | n,0\right)$, the relation between $P$ and $Q$ is given by the following renewal-like equation \cite{carmichael1989photoelectron} 
\begin{eqnarray} 
P_{nm}\left(\tau\right) &=& 
Q_{nm}\left(\tau\right)
+ \kappa \sum_{k=1}^{+\infty} k 
\left( P_{nk-1} \ast 
Q_{km} \right)\left(\tau\right)
\, ,  
\label{RelationgPQ1} 
\end{eqnarray} 
where we wrote $\left( g \ast h \right) \left(\tau\right) \equiv 
\int_0^\tau d\tau_1 g\left(\tau-\tau_1\right) h\left(\tau_1\right)$ the convolution between any two causal functions $g$ and $h$. 
After Laplace-transform, \Eqref{RelationgPQ1} becomes an algebraic equation 
\begin{eqnarray} 
\tilde{P}_{nm}\left(z\right) &=& 
\tilde{Q}_{nm}\left(z\right)
+ \kappa \sum_{k=1}^{+\infty} k 
\tilde{P}_{nk-1}\left(z\right) 
\tilde{Q}_{km}\left(z\right)
\, .  
\label{RelationgPQ2} 
\end{eqnarray} 
\subsection{Average delay-time between two successive photon-emission events}
\label{SecWTD2}
In this section, we derive an analytical expression for the average delay-time between two successive photon-emission events
$\left\langle \tau \right\rangle=\int_0^{+\infty} d\tau \tau w(\tau)$, 
using Eqs.~\eref{EigenmodeDecomposition4}, \eref{WTD4} and \eref{RelationgPQ2}. 
The idea of this derivation is based on the fact that the Laplace-transform of $w(\tau)$
is the moment generating function $\tilde{w}(z)=\int_0^{+\infty} d\tau e^{-z\tau} w(\tau)$ of the WTD.
Its analytical behavior for $z \rightarrow 0$, provides $\tilde{w}(z) \approx 1 - z\left\langle \tau \right\rangle + \cdots$.
Similarly to the theoretical approaches used for computing the mean reaction time of polymer reactions \cite{guerin2012non,guerin2013reactive,PhysRevE.87.032601,condamin2008probing}, we thus perform a systematic expansion \footnote{T. Gu\'{e}rin, private communication.} of \Eqref{RelationgPQ2} for $z \rightarrow 0$, using the following expansion of $\tilde{P}$ and $\tilde{Q}$ 
\begin{eqnarray} 
\tilde{P}_{nm}(z) &\approx & \frac{\pi_n^{\rm{(st)}}}{z} + \tilde{\mathcal{G}}_{nm}(0) + \rm{o}\left(z^0\right)
\, ,  
\label{zExp1} \\
\kappa n \tilde{Q}_{nm}(z) &\approx & \pi_{n|m} - z \tau_{n|m}
+ \rm{o}\left(z\right)
\, , 
\label{zExp2}
\end{eqnarray} 
with the coefficients
\begin{eqnarray} 
\pi_{n|m} &=& \kappa n \int_0^{+\infty} d\tau Q_{nm}(\tau)
\, ,  
\label{zExp3} \\
\tau_{n|m} &=& \kappa n \int_0^{+\infty} d\tau \tau Q_{nm}(\tau)
\, . 
\label{zExp4}
\end{eqnarray} 
\subsubsection{Lowest-order $\rm{o}\left(1/z\right)$}
\label{SecWTD21}
The lowest-order in expanding \Eqref{RelationgPQ2} is of order $\rm{o}\left(1/z\right)$, and provides the following condition
\begin{eqnarray} 
\sum_{n=0}^{+\infty} \pi_{n|m} &=& 1 
\, . 
\label{LowestOrder1}
\end{eqnarray} 
Together with its definition given in \Eqref{zExp3}, $\pi_{n|m}$ can thus be interpreted as a splitting-probability \cite{guerin2012non,condamin2008probing}, namely
it is the probability of first photon-emission event from state $n$ to $n-1$ (integrated in time), given that the state $m$ was realized at time $0$. 
The notion of splitting-probability enables, with the definition in \Eqref{zExp4}, to interpret $\tau_{n|m}$ as the average-time of first photon-emission event from state $n$ to $n-1$, given that the state $m$ was realized at time $0$.
We remark that formally, the average delay-time $\left\langle \tau \right\rangle$ between two successive photon emission events can be obtained from the times $\tau_{n|m}$ as
\begin{eqnarray} 
\left\langle \tau \right\rangle &=& \frac{1}{\left\langle n \right\rangle}
\sum_{n,m=0}^{+\infty} \tau_{n|m} \left(m+1\right) \pi_{m+1}^{\rm{(st)}}
\, . 
\label{LowestOrder2}
\end{eqnarray} 
\subsubsection{Order $\rm{o}\left(z^0\right)$}
\label{SecWTD22}
The next order in expanding \Eqref{RelationgPQ2} is of order $\rm{o}\left(z^0\right)$. 
It provides, using \Eqref{LowestOrder2}, the following system of equations
\begin{eqnarray} 
\xi_n - \kappa^2 n \pi_{n}^{\rm{(st)}} \left\langle n \right\rangle  \left\langle \tau \right\rangle &=& \kappa^2 n \sum_{m=1}^{+\infty}  \tilde{\mathcal{G}}_{nm-1}(0) m \pi_{m}^{\rm{(st)}}
\, \nonumber \\
&-& \kappa n \sum_{m=1}^{+\infty} \tilde{\mathcal{G}}_{nm-1}(0) \xi_{m}
\, ,
\label{ZeroOrder1}
\end{eqnarray} 
with the unknowns $\xi_n = \kappa \sum_{m=0}^{+\infty} \pi_{n|m} \left(m+1\right) \pi_{m+1}^{\rm{(st)}}$ and $\left\langle \tau \right\rangle$.
Using \Eqref{LowestOrder1}, we find that $\sum_{n=0}^{+\infty} \xi_n = \kappa \left\langle n \right\rangle$, and get the general solution of \Eqref{ZeroOrder1} 
\begin{eqnarray} 
\xi_n &=& \kappa n \pi_{n}^{\rm{(st)}}
\, \label{ZeroOrder2} \\
\left\langle \tau \right\rangle &=& \frac{1}{\kappa \left\langle n \right\rangle} \equiv \frac{1}{\Gamma_\gamma^{(\rm{st})}}
\, ,
\label{ZeroOrder3}
\end{eqnarray} 
Eqs.~\eref{ZeroOrder2} and \eref{ZeroOrder3} are the main results of this section.
\Eqref{ZeroOrder3} states that the stationary probability per unit of time of emitting a photon out of cavity is the inverse average delay-time between two successive photon emission events. 
This relation is reminiscent of Kac's lemma \cite{kac1947notion,aldous2002reversible}, and holds generally for ergodic systems.
\subsection{Relation between the SCF and WTD}
\label{SecWTD3}
\Eqref{RelationgPQ1} provides a relation between the $P$ and $Q$ distributions. 
However, it does not imply a simple relation between $S(\tau)$ (see \Eqref{WTD3}) and $w(\tau)$ (see \Eqref{WTD4}), except in the case of a two-level atom or molecule (see \Secref{AnSym2}).
In general, the complete distribution function $P(\tau |0)$ is formally obtained as the sum of all intermediate histories containing "no photon emission", plus "one-photon emission event", plus "two-photon emission events", plus etc $\cdots$ between the two photon-emission events in the time-interval $\left\rbrack 0,\tau \right\lbrack$.
This can be written mathematically as the following series
\begin{eqnarray}
P(\tau |0) &=&
Q(\tau |0)
+
\int_{0}^{\tau} d\tau_1 Q(\tau,\tau_1 | 0)
\nonumber \, \\
&+&
\int_{0}^{\tau} d\tau_1 \int_{\tau_1}^{\tau} d\tau_2 
Q(\tau,\tau_2,\tau_1 | 0)
+
\cdots  \,  ,
\nonumber \, \\
\label{RelSCFWTD1} 
\end{eqnarray}
with $Q(\tau,\tau_1 | 0)$ the exclusive probability distribution that one photon is emitted at time $\tau$, and another one is emitted at time $\tau_1$, knowing that one photon-emission event occurred at time $0$.
A similar definition holds for $Q(\tau,\tau_2,\tau_1 | 0)$ that contains an additional photon-emission event at time $\tau_2$.
If the stochastic process of emitting photons out of the cavity is Markovian, then $Q(\tau,\tau_1 | 0)\equiv Q(\tau| \tau_1)Q(\tau_1| 0)$ and $Q(\tau,\tau_2,\tau_1 | 0)\equiv Q(\tau|\tau_2)Q(\tau_2| \tau_1)Q(\tau_1| 0)$.
\Eqref{RelSCFWTD1} thus reduces in this case to
\begin{eqnarray}
P(\tau |0) &=&
Q(\tau |0)
+
\int_{0}^{\tau} d\tau_1 Q(\tau| \tau_1)Q(\tau_1| 0)
\nonumber \, \\
&+&
\int_{0}^{\tau} d\tau_1 \int_{\tau_1}^{\tau} d\tau_2 
Q(\tau| \tau_2)Q(\tau_2| \tau_1)Q(\tau_1| 0)
\nonumber \, \\
&+&
\cdots  \,  ,
\nonumber \, \\
\label{RelSCFWTD2} 
\end{eqnarray}
This series can be resummated into an integral equation
\begin{eqnarray}
P(\tau |0) &=&
Q(\tau |0)
+
\int_{0}^{\tau} d\tau_1 P(\tau|\tau_1)Q(\tau_1| 0)
\, . \nonumber \\
\label{RelSCFWTD3} 
\end{eqnarray}
Using the definition of $S(\tau)$ (see \Eqref{WTD1}) and $w(\tau)$ (see \Eqref{WTD2}), \textit{if the stochastic process associated to the emission of photons out of the cavity is both Markovian and stationary}, we finally obtain  
\begin{eqnarray}
S(\tau) &=& w(\tau) + 
\left( S \ast w \right)(\tau)
\label{RelSCFWTD4} \, . 
\end{eqnarray}
\Eqref{RelSCFWTD4} can be solved by Laplace-transform, thus providing a simple algebraic link between $\tilde{S}(z)$ and $\tilde{w}(z)$, the respective Laplace transforms of $S(\tau)$ and $w(\tau)$
\begin{eqnarray}
\tilde{w}(z) &=& \frac{1}{1+\tilde{S}^{-1}(z)}
\label{RelSCFWTD5} \, . 
\end{eqnarray}
%

%
\section{Analytical results for a symmetric junction}
\label{AnSym}
\subsection{Analytical solutions of the RE}
\label{AnSym1}
%
\begin{figure}[t!]
\centering
\includegraphics[width=1.0\linewidth]{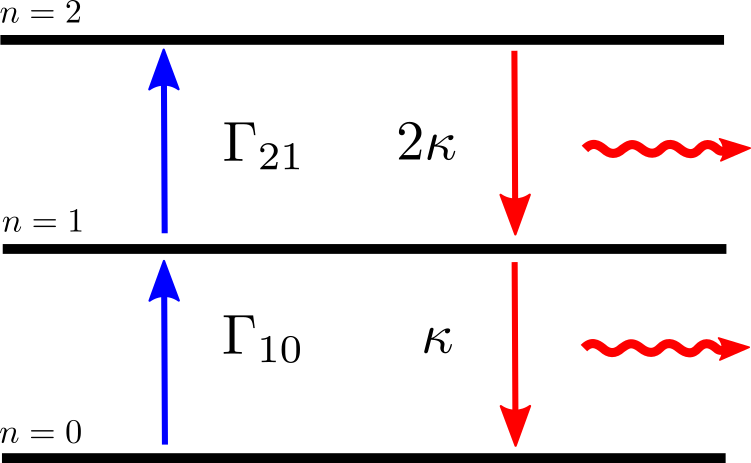}
\caption{
Graphical representation of the truncated rate equation \Eqref{Analytics_1} giving the cavity-mode populations $\pi_n(t)$, in the case of a symmetric junction close to the first inelastic threshold for photon emission.
Red (blue) downward (upward) arrows stand for dominant (subdominant) transition rates induced by cavity-damping (inelastic tunneling of electrons) .
}
\label{fig:Fig2}
\end{figure}
%
In this section, we restrict to the case of electron-hole symmetric junctions, and bias voltages close to the first inelastic threshold for photon emission $\mu_L=-\mu_R=eV/2 \approx \hbar\omega_c$.
In this regime, the dominant transitions rates are represented in \Figref{fig:Fig2}, as red downward arrows for cavity-damping rates ($\kappa$ and $2\kappa$) and blue upward arrows for inelastic tunneling rates of single electrons across the STM junction ($\Gamma_{10}$ and $\Gamma_{21}$).
\Eqref{RE_3} then reduces to an approximate RE involving  a vector of populations of lower dimension $\underline{\pi}^{(r)}(t)=\lbrack \pi_0(t),\pi_1(t),\pi_2(t) \rbrack$ (see \Figref{fig:Fig2}) 
\begin{eqnarray}
\dot{\underline{\pi}}^{(r)}(t) &=& \mathbf{\Gamma}^{(r)}\underline{\pi}^{(r)}(t)
\label{Analytics_1} \, , \\
\mathbf{\Gamma}^{(r)} &\approx& \begin{bmatrix}
-\Gamma_{10} & \kappa & 0\\
\Gamma_{10} & -(\kappa + \Gamma_{21}) & 2\kappa \\
0 & \Gamma_{21} & -2\kappa
\end{bmatrix} \label{Analytics_2}  \, , 
\end{eqnarray}
Both left, right eigenvectors and eigenvalues of the reduced $\mathbf{\Gamma}^{(r)}$-matrix can be found analytically, although the resulting expression for computing the SCF and WTD are lengthy.
We thus resort on the limit $\Gamma \ll \kappa$, for which the following approximation of the non-zero eigenvalues $\lambda_\pm$ is found 
\begin{eqnarray}
\lambda_+ &\approx & - \left( \kappa + \Gamma_{10} - \Gamma_{21}\right)
\approx - \kappa
\label{Analytics_3} \, , \\
\lambda_- &\approx & - 2\left( \kappa + \Gamma_{21}\right)
\approx - 2\kappa
\label{Analytics_4} \, ,
\end{eqnarray}
as well as the stationary populations of the cavity-mode
\begin{eqnarray}
\pi_1^{\rm{st}} &\approx & \frac{\Gamma_{10}}{\kappa}
\label{Analytics_5} \, , \\ 
\pi_2^{\rm{st}} &\approx & \frac{\Gamma_{10}\Gamma_{21}}{2\kappa^2}
\label{Analytics_6} \, , \\
\pi_0^{\rm{st}} &=& 1 - \pi_1^{\rm{st}} - \pi_2^{\rm{st}}
\label{Analytics_6bis} \, .
\end{eqnarray}
The corresponding approximate solution of \Eqref{Analytics_1} is obtained in this limit, for any initial condition $\underline{\pi}^{(r)}(0)=\lbrack \pi_0(0),\pi_1(0),\pi_2(0)\rbrack$ as 
\begin{eqnarray}
\pi_1(t) &\approx & \pi_1^{\rm{st}} + \left\lbrace
\pi_1(0)+2\pi_2(0) - \left\lbrack
\pi_1^{\rm{st}} + 2\pi_2^{\rm{st}}
\right\rbrack
\right\rbrace e^{-\kappa t}
\nonumber \\
&+&
2 \left\lbrace
\pi_2^{\rm{st}} - \pi_2(0)
\right\rbrace e^{-2\kappa t}
\label{Analytics_7} \, , \\ 
\pi_2(t) &\approx & \pi_2^{\rm{st}}
+
\left\lbrace
\pi_2(0) - \pi_2^{\rm{st}}
\right\rbrace e^{-2\kappa t}
\label{Analytics_8} \, ,\\
\pi_0(t) &=&  1 - \pi_1(t) - \pi_2(t)
\label{Analytics_8bis} \, .
\end{eqnarray}
\subsection{Analytical result for the SCF}
\label{AnSym02}
%
%
\begin{figure}[t!]
\includegraphics[width=1.0\linewidth]{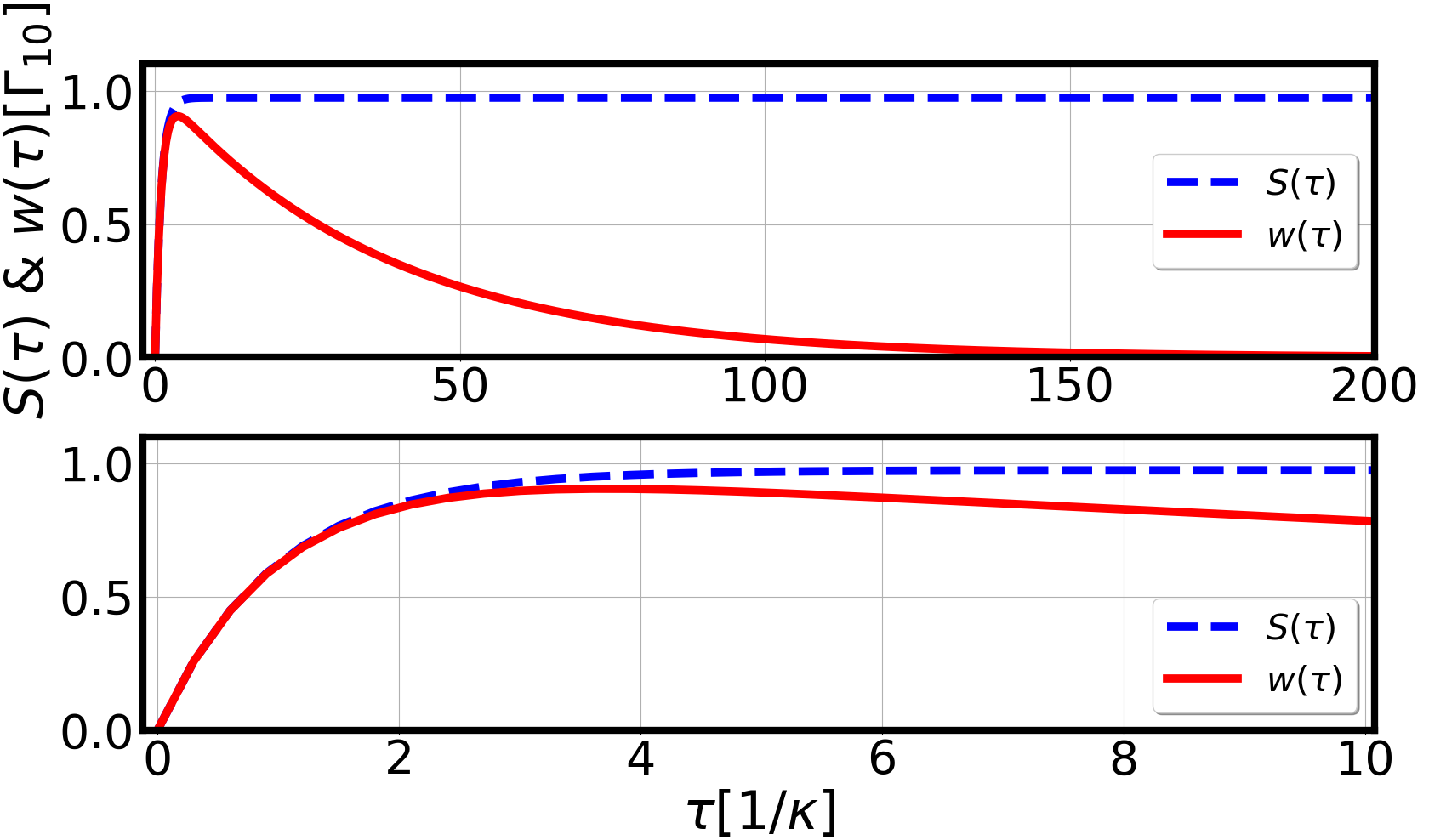}
\caption{
Time-dependence of $S(\tau)=\kappa \left\langle n \right\rangle g^{(2)}(\tau)$ (dashed-blue curves) obtained from \Eqref{SymWTD1}, and $w(\tau)$ (plain red curves) obtained from \Eqref{SymWTD3}, both expressed in units of $\Gamma_{10}$.
Upper panel: "Long" time-scale $\tau \geq \tau_{\rm{m}}$ in units of $1/\kappa$, with $\tau_{\rm{m}}$ given by \Eqref{SymWTD6}.
Lower panel: Same curve but on "short" time-scale $\tau \leq \tau_{\rm{m}}$.
Parameters are : $\kappa=k_BT/\hbar=0.1\omega_c$, $\Gamma_L=\Gamma_R=\Gamma=0.1\kappa$, $\mu_L=-\mu_R=eV/2=\hbar\omega_c$, $\lambda=\sqrt{2}$, and $\tilde{\varepsilon}_0=0$.
}
\label{fig:Fig3}
\end{figure}
%
Using the results of the previous \Secref{AnSym1}, Eqs.~\eref{Glauber_2} and \eref{Glauber_4}, we obtain the following approximate expressions %
\begin{eqnarray}
G^{\rm{st}}_1 &\approx & \kappa \left( \pi_1^{\rm{st}} + 2\pi_2^{\rm{st}}  \right) = \frac{\Gamma_{10}(\kappa + \Gamma_{21})}{\kappa} \approx \Gamma_{10}
\label{Analytics_9} \, , \\ 
G^{\rm{st}}_2(\tau) &\approx & \kappa^2 \left\lbrace 
\left( \pi_1^{\rm{st}} + 2\pi_2^{\rm{st}} \right)^2
\left( 1 - e^{-\kappa \tau} \right) +
2\pi_2^{\rm{st}} e^{-\kappa \tau}
\right\rbrace
\, , \nonumber \\
\label{Analytics_10}
\end{eqnarray}
from which the SCF of emitted photons is obtained as 
\begin{eqnarray}
g^{(2)}(\tau) &\approx & 1 + e^{-\kappa \tau} \left( 
g^{(2)}(0) - 1
\right)
\label{Analytics_11} \, , \\
g^{(2)}(0) &=& \frac{\langle n(n-1) \rangle}{\langle n \rangle^2}
= \frac{2\pi_2^{\rm{st}}}{\left( \pi_1^{\rm{st}} + 2\pi_2^{\rm{st}} \right)^2}
\approx 
\frac{\Gamma_{21}}{\Gamma_{10}}
\label{Analytics_12} \, . 
\end{eqnarray}
Eqs.~\eref{Analytics_12} and \eref{Analytics_11} are the main results of this section.
They recover and confirm independently the numerical results of Ref.~\onlinecite{PhysRevLett.123.246601}, that $g^{(2)}(\tau)$ decays exponentially in time with the cavity-damping rate $\kappa$.
We find a  
\begin{enumerate}
\item 
Bunching behavior ($g^{(2)}(\tau) \leq g^{(2)}(0) \approx 2$) for $\lambda \ll 1$,
\item
Complete antibunching ($g^{(2)}(\tau) \geq g^{(2)}(0)=0$) for $\lambda = \sqrt{2}\approx 1.41$, 
\item
Poissonian behavior ($g^{(2)}(\tau) = g^{(2)}(0) = 1$) for $\lambda = \sqrt{2-\sqrt{2}}\approx 0.77$.
\end{enumerate}
\subsection{Special case of photon antibunching ($\lambda=\sqrt{2}$)}
\label{AnSym2}
\subsubsection{Occupation probability of the cavity mode}
\label{Antibunching1}
In this section, we focus on the case of plasmon-molecule coupling $\lambda=\sqrt{2}$, for which perfect antibunching is obtained
in \Eqref{Analytics_11}.
Due to the vanishing of the Franck-Condon overlap between the states $n=1$ and $n=2$ of the displaced cavity-mode, we have that $\Gamma_{21}=0$, so that the excited cavity-state $n=2$ in \Figref{fig:Fig2} cannot be reached. 
\Eqref{Analytics_1} describing the physical problem of light-emission, thus simplifies to a problem involving only two cavity states $n=0$ and $n=1$ 
\begin{eqnarray}
\dot{\pi}_1(t) &\approx & \Gamma_{10} \pi_0(t) - \kappa \pi_1(t) 
\label{SimpleEq1} \, ,    \\
\pi_0(t) &=& 1 - \pi_1(t)
\label{SimpleEq2} \, . 
\end{eqnarray}
Its solution for the initial condition $\pi_1(0)=0$ (initially empty cavity) is given by
\begin{eqnarray}
\pi_1(t) &=&
\pi^{\rm{st}}_1 \left\lbrace 1 - e^{-\kappa_t t} \right\rbrace
\label{SimpleEq3} \, ,  \\
&\approx & \frac{\Gamma_{10}}{\kappa} \left( 1 - e^{-\kappa t} \right)
\label{SimpleEq4} \, ,
\end{eqnarray}
with $\kappa_t = \kappa + \Gamma_{10}$, and $\pi^{\rm{st}}_1=\Gamma_{10}/\kappa_t$.
In the rest of this supplementary material, the approximate sign $\approx$, as written in \Eqref{SimpleEq4}, will mean that the limit $\kappa \gg \Gamma_{10}$ is taken.
We note that the stationary occupation of the cavity-mode is reached for time-scales $t \gg 1/\kappa$ and is given by $\pi^{\rm{st}}_1 \approx \Gamma_{10}/\kappa$, which comes from the balance between the rate at which the cavity is populated (with a rate given by the inelastic electronic tunneling rate $\Gamma_{10}$) and the rate at which the energy is dissipated (with cavity dissipation-rate $\kappa$).
At shorter time scales $0 \leq t \ll 1/\kappa$, the population of the cavity modes is given by $\pi_1(t)\approx \Gamma_{10} t$, and thus grows linearly with the inelastic tunneling rate $\Gamma_{10}$.
This reflects the fact that starting with an empty cavity $n=0$, a new electron has to tunnel into the dot so that a cavity photon is emitted to reach the occupancy $n=1$.
\subsubsection{SCF for the emitted photons}
\label{Antibunching2}
Using the results of the previous \Secref{Antibunching1}, we obtain the first-order correlation function  
\begin{eqnarray}
G^{\rm{st}}_1 = \kappa \pi^{\rm{st}}_1 = \frac{\kappa}{\kappa_t} \Gamma_{10} \approx \Gamma_{10}
\label{SimpleEq5} \, .
\end{eqnarray}
The probability that a photon is emitted out-of-cavity between times $t_1$ and $t_1+dt_1$, is thus proportional to the inelastic electronic tunneling rate $\Gamma_{10}$.
In other words, an electron has to tunnel onto the dot and emit a cavity-photon that further decays out of the cavity.
We obtain in the same way the SCF in this regime
\begin{eqnarray}
G^{\rm{st}}_2(\tau) &=& \kappa^2 P\left( 1 \tau | 0 0 \right)
\pi^{\rm{st}}_1
\nonumber \\
&=& \left(\frac{\kappa\Gamma_{10}}{\kappa_t}\right)^2
\left( 1 - e^{-\kappa_t\tau} \right)
\approx  \Gamma_{10}^2 \left( 1 - e^{-\kappa\tau} \right)
\nonumber  \, , \\
\label{SimpleEq6} \\
g^{(2)}(\tau) &=& 1 - e^{-\kappa_t\tau} \approx  1 - e^{-\kappa\tau} 
 \, . \label{SimpleEq6bis}
\end{eqnarray}
After the first photon has been emitted between time $t_1$ and $t_1+dt_1$, the cavity mode gets back to its ground-state $n=0$ and an additional electron tunneling event is needed to repopulated the cavity mode to $n=1$ that fast decays by emission of a second photon out-of-cavity between times $t_2$ and $t_2+dt_2$.
This refilling of the cavity-mode is encoded by the time-evolution of  
$\pi_1(\tau)$ in \Eqref{SimpleEq3}.
We obtain at large times $\tau \gg 1/\kappa$ that $G^{\rm{st}}_2(\tau)\approx \Gamma_{10}^2 \equiv \left( G^{\rm{st}}_1 \right)^2$, hence the two photon emission events become independent one from each other.
At short times $\tau \ll 1/\kappa$, we find $G^{\rm{st}}_2(\tau)\approx \Gamma_{10}^2 \kappa \tau$, and the initial slope $\dot{G}^{\rm{st}}_2(0)=\Gamma_{10}^2 \kappa$.
This is consistent with the argument that two inelastic electronic tunneling (rare) events are necessary for two photons to be emitted, thus explaining the $\Gamma_{10}^2$ factor, while being not incompatible with the fact that the correlation function itself $g^{(2)}(\tau) = G^{\rm{st}}_2(\tau)/\left(G^{\rm{st}}_1\right)^2$ decays with the damping rate $\kappa$ (frequent decay events) in \Eqref{SimpleEq6bis}.
Finally, we remark that \Eqref{SimpleEq6bis} is consistent with \Eqref{Analytics_11} for the case $\lambda = \sqrt{2}$.
\subsubsection{WTD of emitted photons}
\label{AnSym3}
%
\begin{figure}[t!]
\includegraphics[width=1.0\linewidth]{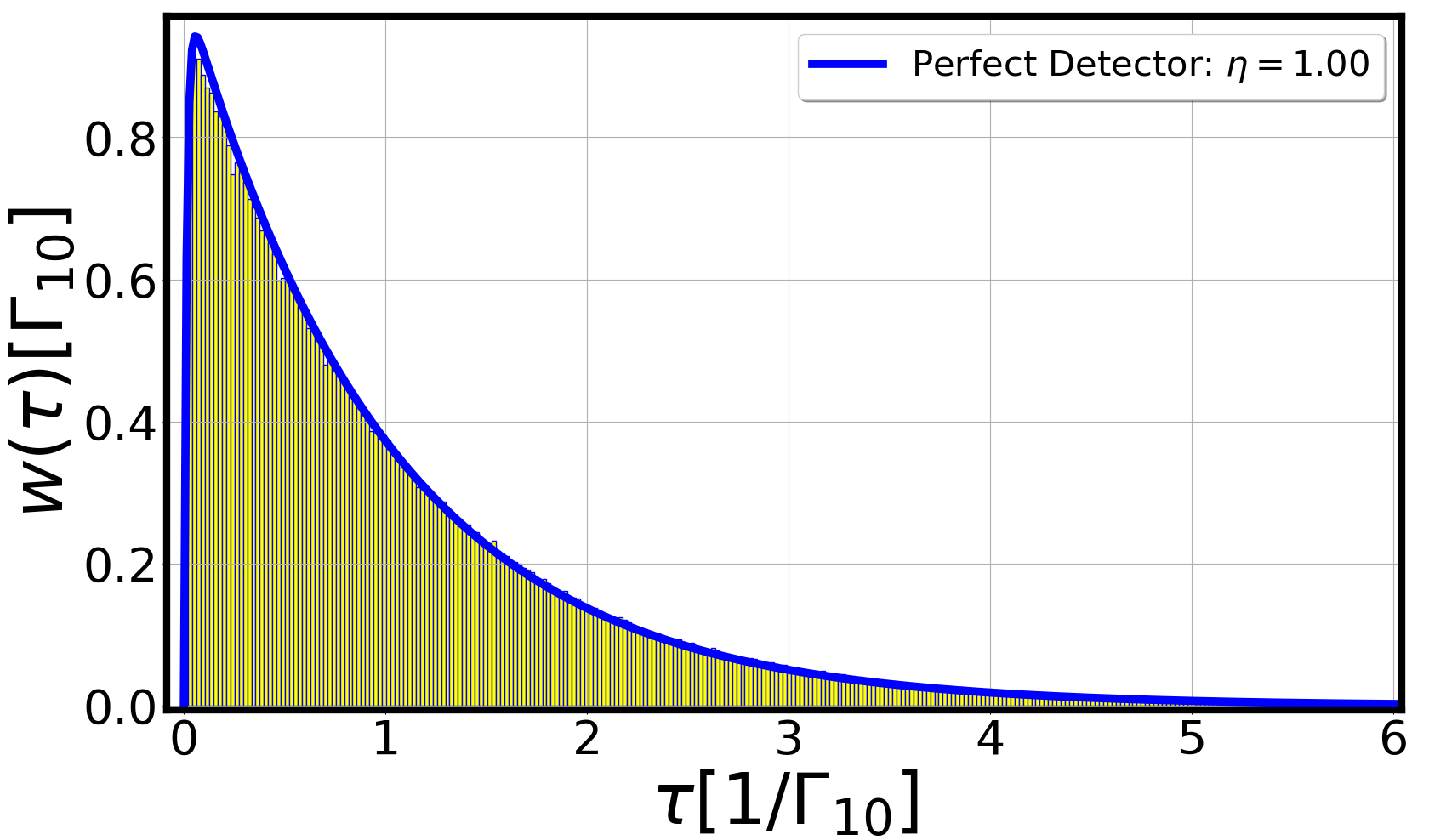}
\caption{
Time-dependence of $w(\tau)$ (histogram) obtained from the Monte Carlo numerical calculation, expressed in units of $\Gamma_{10}$.
The plain blue curve is the corresponding analytical result of \Eqref{SymWTD3}.
Parameters are those of \Figref{fig:Fig3}.
}
\label{fig:Fig4}
\end{figure}
For completeness, we present in \Figref{fig:Fig4} the WTD $w(\tau)$ (histogram), as obtained from the output of the numerical Monte Carlo calculation.
We remark that the histogram is well approximated by the 
analytical distribution (see plain blue curve) shown by \Eqref{SymWTD3}.
%
%
%
%
%
%
%
%
%
In this section, we compute $w(\tau)$ analytically.
We first remark that in the regime of \Secref{AnSym2}, \Eqref{RelationgPQ1} can be rewritten, after restricting to the available states of the cavity $n=0,1$, in the form of \Eqref{RelSCFWTD4}.
In other words, once $g^{(2)}(\tau)$ is known, $w(\tau)$ can be obtained by solving the integral \Eqref{RelSCFWTD4}.
We pursue further by rewriting explicitly $S(\tau)=\kappa \left\langle n \right\rangle g^{(2)}(\tau)$ using \Eqref{SimpleEq6bis}, and performing its Laplace transform $\tilde{S}(z)$
\begin{eqnarray}
S(\tau) &=& \frac{\kappa \Gamma_{10}}{\kappa_t}
\left\lbrace
1 - e^{-\kappa_t\tau}
\right\rbrace
\label{SymWTD1} \, ,
\\
\tilde{S}(z) &=&
\frac{\kappa \Gamma_{10}}{z\left( z + \kappa_t \right)} 
\label{SymWTD2} \, .
\end{eqnarray}
Incorporating \Eqref{SymWTD2} into the expression of $\tilde{w}(z)$ in terms of $\tilde{S}(z)$ (see \Eqref{RelSCFWTD5}), we find after inverse Laplace transform
\begin{eqnarray}
w(\tau) &=&  \frac{\kappa \Gamma_{10}}{\kappa - \Gamma_{10}} \left\lbrace
e^{-\Gamma_{10}\tau}
-
e^{-\kappa\tau}
\right\rbrace
\label{SymWTD3} \, , \\
& \approx & \kappa \Gamma_{10} \tau \xrightarrow[\tau \to 0^+]{} 0^+
\label{SymWTD4} \, , \\
& \approx & 
\Gamma_{10} e^{-\Gamma_{10}\tau}
\xrightarrow[\tau \to +\infty]{} 0^+
\label{SymWTD5} \, .
\end{eqnarray}
\Eqref{SymWTD3} is the main result of this section, showing that $w(\tau)$ is the sum of two exponentials, one with relaxation rate $\Gamma_{10}$ and the other with relaxation rate $\kappa$. 
We show on \Figref{fig:Fig3}, the time-evolution of $S(\tau)$ (dashed blue curves) given by \Eqref{SymWTD1} and $w(\tau)$ (plain red curves) given by \Eqref{SymWTD3}.
At "short" times ($\tau \ll 1/\kappa$), $w(\tau)$ vanishes linearly with $\tau$, with a rate proportional to the inelastic electronic tunneling rate $\Gamma_{10}$ (see \Eqref{SymWTD4} and \Figref{fig:Fig3}-lower panel).
The vanishing of $w(\tau)$ at coincidence time $\tau=0$ is a signature of antibunching of emitted photons, while the linear slope with $\Gamma_{10}$ translates the fact that after one photon has been emitted, one needs to wait some delay time for refilling the cavity mode (by charging or discharging the dot) and have another photon emitted.
In the opposite limit of "long" times ($\tau \gg 1/\kappa$), $w(\tau)$ vanishes exponentially in time with the rate $\Gamma_{10}$ (see \Eqref{SymWTD5} and \Figref{fig:Fig3}-upper panel).
This reflects both that $w(\tau)$ is a normalized distribution function $\left\lbrack \int_0^{+\infty} d\tau w(\tau) = 1 \right\rbrack$, and that it becomes very unlikely that after one photon has been emitted initially, another photon is not emitted after a delay-time much larger than the cavity damping-time.
It is interesting to notice that $w(\tau)$ is a non-monotonous function  of time, reaching a maximum at time $\tau_{\rm{m}}\approx 1/\kappa$ given by
\begin{eqnarray}
\tau_{\rm{m}} &=& \frac{1}{\kappa-\Gamma_{10}}
\ln \left(
\frac{\kappa}{\Gamma_{10}}
\right)
\label{SymWTD6} \, .
\end{eqnarray}
Finally, we obtain from \Eqref{SymWTD3}, the average delay-time $\left\langle \tau \right\rangle$ between two photon emission events
\begin{eqnarray}
\left\langle \tau \right\rangle &=& \int_0^{+\infty} d\tau \tau w(\tau)
= \frac{1}{\Gamma_{10}} + \frac{1}{\kappa} \approx  \frac{1}{\Gamma_{10}}
\label{SymWTD7} \, .
\end{eqnarray}
\Eqref{SymWTD7} recovers the general expression given by \Eqref{ZeroOrder3}.
We thus conclude that $S(\tau)$ decays exponentially in time with the total dissipation rate $\kappa + \Gamma_{10}$, and thus 
with a relaxation time $1/\left( \kappa + \Gamma_{10} \right)\approx 1/\kappa$, as reported in Ref.~\onlinecite{PhysRevLett.123.246601}.
In contrast to $S(\tau)$, $w(\tau)$ exhibits a non-monotonous behavior with time, from which the average delay-time between two photon emission events $\left\langle \tau \right\rangle$ is obtained as the sum of the inverse relaxation time $1/\Gamma_{10}$ associated to inelastic electronic tunneling plus the inverse relaxation time of cavity photon-decay $1/\kappa$. 
In summary, $S(\tau)$ and $w(\tau)$ do not contain the same statistical information, each exhibiting different characteristic time-scales characterizing a different aspect of the photon-emission statistics.
Those two time-scales are not inconsistent one with each other but constitute complementary statistical indicators. 
Only in the short-time regime $\tau \ll \tau_{\rm{m}}$, the behavior in time of $w(\tau)$ and $S(\tau)$ coincide (being related to the same antibunching mechanism), while at large time $\tau \gg \tau_{\rm{m}}$ they differ in a significant manner, since $w(\tau)$ goes exponentially to zero while $S(\tau)$ converges to $\Gamma_{10}$ (see \Figref{fig:Fig3}).
%

%
%
%
%
%
%
%
For completeness, we present in \Figref{fig:Fig4} the WTD $w(\tau)$ (histogram), as obtained from the output of the numerical Monte Carlo calculation.
We remark that the histogram is well approximated by the 
analytical distribution (see plain blue curve) shown by \Eqref{SymWTD3}.
%

%
\subsection{Special case at the crossover point ($\lambda=\sqrt{2-\sqrt{2}}$)}
\label{Crossover_Point}
%
%
\begin{figure}[t!]
\includegraphics[width=1.0\linewidth]{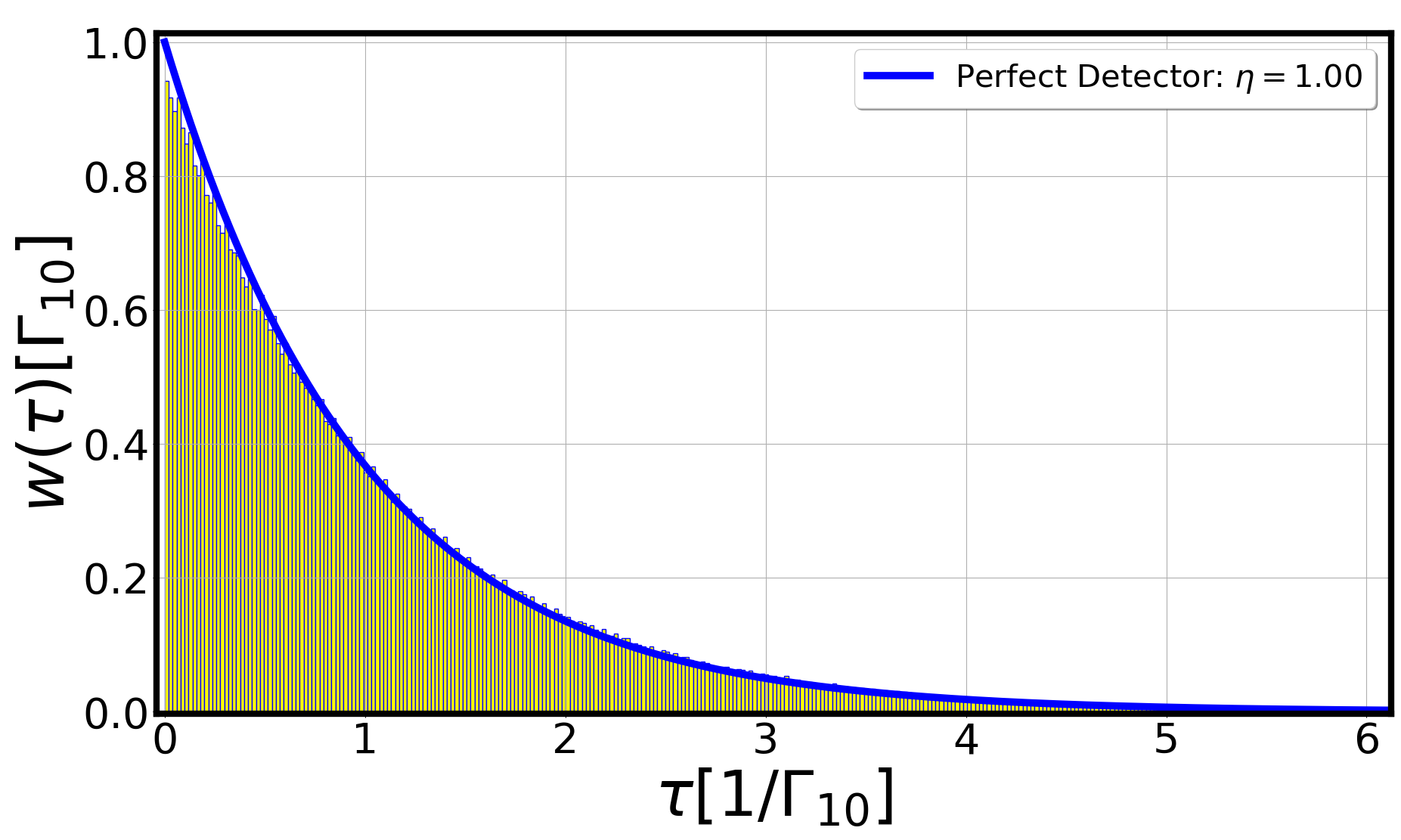}
\caption{
Time-dependence of $w(\tau)$ (histogram) obtained from the Monte Carlo numerical calculation, expressed in units of $\Gamma_{10}$.
The plain blue curve is given by the approximated analytical formula $w(\tau) \approx \Gamma_{10} e^{-\Gamma_{10}\tau}$.
Parameters are those of \Figref{fig:Fig3}, but with a modified coupling strength $\lambda=\sqrt{2-\sqrt{2}}$.
}
\label{fig:Fig4bis}
\end{figure}
\textcolor{black}{
Finally, we present in \Figref{fig:Fig4bis} the WTD $w(\tau)$ (histogram), obtained from the numerical Monte Carlo calculation, at the crossover point between the bunching and antibunching regime ($\lambda=\sqrt{2-\sqrt{2}}\approx 0.77$).
For this particular value of the coupling strength, we have shown both numerically (see Fig.2 of the paper, cyan upper triangles) and analytically
(see \Secref{AnSym02}) that the second-order correlation function is constant in time and thus presents a Poissonian behavior, namely that $g^{(2)}(\tau) = g^{(2)}(0) = 1$.
We thus expect the WTD in this regime to be exponentially decreasing in time.
We show that this is indeed the case (see histogram in \Figref{fig:Fig4bis}), on the whole time-window.
The explicit time-dependence of the WTD is provided by the approximated analytical formula $w(\tau) \approx \Gamma_{10} e^{-\Gamma_{10}\tau}$ (see plain blue curve), with the decay rate provided by $\Gamma_{10}$ (evaluated at $\lambda=\sqrt{2-\sqrt{2}}$).
}
%

%
%
%
%
%
%
%
In this section, we model the non-perfect detection efficiency of photodetectors collecting the photons emitted out of the cavity. 
We assign a probability $\eta$ for a photon to be detected by the photodetector, namely $\eta$ is the experimental detection-yield.
A perfect detector efficiency means $\eta=1$.
For computing how $g^{(2)}(\tau)$ and $w(\tau)$ are impacted by this detector efficiency, we need to modify the definitions of \Secref{SCF_General1} and \Secref{SecWTD1}
\begin{enumerate}
\item
$G_1(t_1)dt_1 \equiv$ probability to "emit a photon and detect it by the apparatus" in the time-interval $\lbrack t_1, t_1 + dt_1 \rbrack$.
\item
$P\left( t_1, t_2\right)dt_1 dt_2 \equiv$ joint-probability that "a first photon is emitted and detected by the apparatus" in the time interval $\lbrack t_1, t_1 + dt_1 \rbrack$ and that a "second photon is emitted and detected by the apparatus" in the time interval $\lbrack t_2, t_2 + dt_2 \rbrack$.
\item
$Q\left( t_1, t_2\right)dt_1 dt_2 \equiv$ joint-probability that "a first photon is emitted and detected by the apparatus" in the time interval $\lbrack t_1, t_1 + dt_1 \rbrack$ and that a "second photon is emitted and detected by the apparatus" in the time interval $\lbrack t_2, t_2 + dt_2 \rbrack$, without any photon emitted in the time interval $\rbrack t_1 + dt_1, t_2 \lbrack$.
\end{enumerate}
%
%
\begin{figure}[t!]
\includegraphics[width=1.0\linewidth]{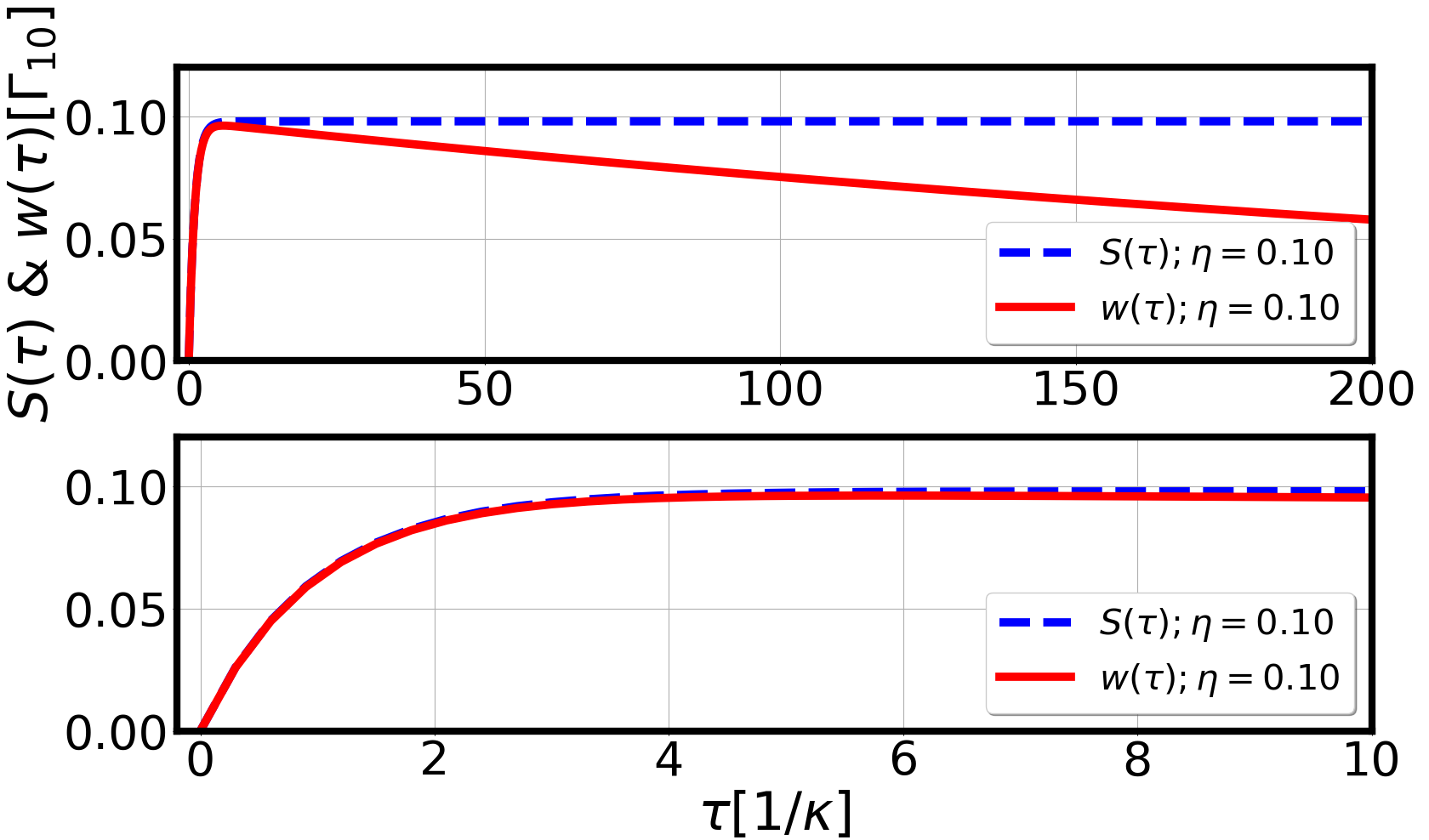}
\caption{
Same curves as for \Figref{fig:Fig3}, but for a low detection yield $\eta=0.1$ as frequently obtained in experimental situations.
}
\label{fig:Fig5}
\end{figure}
%
%
The role of a non-perfect detection-yield is taken into account in the Monte Carlo simulations by an additional random process with probability $\eta$, that is independent of the photon-emission events. 
Due to the above mentioned statistical independence between the photon-emission and photon-detection events, the previously defined probability distributions factorize
\begin{eqnarray}
G_1(t_1) &=& \eta \underline{G}_1(t_1)
\label{P1Det} \, , \\
P(t_1, t_2) &=& \eta^2 \underline{P}(t_1, t_2)
\label{P2Det} \, ,
\end{eqnarray}
with the underline symbols corresponding to quantities evaluated at perfect detection efficiency ($\eta=1$).
Those equations imply a modification of \Eqref{g2_1} and \Eqref{WTD1} as
\begin{eqnarray}
S(\tau) &=& P(\tau | 0) = \eta \underline{S}(\tau)
\label{PS1} \, , \\
G^{\rm{st}}_2(\tau) &=& P( \tau , 0)
= \eta^2 \underline{G}^{\rm{st}}_2(\tau)
\label{PS2} \, , \\
g^{(2)}(\tau) &=& \frac{P( \tau | 0)}{G^{\rm{st}}_1(0)} 
\nonumber \\
&=& \frac{\eta \underline{S}(\tau)}{\eta \underline{G}^{\rm{st}}_1}
= \underline{g}_2(\tau)
\label{PS3} \, , 
\end{eqnarray}
so that the unnormalized SCF $G^{\rm{st}}_2(\tau)$ is modified by the non perfect photon-detection quantum-yield, while the normalized SCF $g^{(2)}(\tau)$ is not, since in the former, the $\eta^2$ factors simplify in the numerator and denominator. 
Using the definition of $w(\tau) = Q(\tau | 0)$, \Eqref{RelSCFWTD4} and \Eqref{PS1}, we find that in the regime of photon antibunching ($\lambda=\sqrt{2}$) of \Secref{AnSym2} 
\begin{eqnarray}
\eta \underline{S}(\tau) &=& w(\tau) + 
\eta \left( w \ast \underline{S} \right)(\tau)
\label{PS4} \, , 
\end{eqnarray}
which provides after Laplace transform
\begin{eqnarray}
\tilde{w}(z) &=& \frac{1}{1+\left\lbrace 
\eta\underline{\tilde{S}}(z)\right\rbrace^{-1}}
\label{PS5} \, . 
\end{eqnarray}
\Eqref{PS5} generalizes \Eqref{RelSCFWTD5} to the case of a non-perfect detection yield.
The final WTD is obtained as 
\begin{eqnarray}
w(\tau) &=&  \frac{\eta\kappa \Gamma_{10}}{\kappa_d}
\left\lbrace
e^{-\frac{\left(\kappa_t - \kappa_d\right)\tau}{2}}
-
e^{-\frac{\left(\kappa_t + \kappa_d\right)\tau}{2}}
\right\rbrace
\label{PS6} \, , 
\end{eqnarray}
with $\kappa_d=\sqrt{\kappa_t^2 - 4 \eta\kappa \Gamma_{10}}$.
The corresponding average delay-time $\left\langle \tau \right\rangle $ between two photon emission events is modified to
\begin{eqnarray}
\left\langle \tau \right\rangle &=& \frac{1}{\eta}
\left\lbrace
\frac{1}{\Gamma_{10}} + \frac{1}{\kappa}
\right\rbrace
\label{PS7} \, .
\end{eqnarray}
It is interesting to notice that, in contrast to $g_2(\tau)$, $w(\tau)$ depends significantly on the non-perfect detector efficiency $\eta$. 
\Eqref{PS6} shows indeed that the initial slope of $w(\tau)$ is decreased upon decreasing efficiency, while the rate of its exponential decrease at long times is also modified.
Finally, $\left\langle \tau \right\rangle$ increases towards longer times with a reduction of $\eta$ in \Eqref{PS7}.
This exemplifies the necessity to properly take into account the detector efficiency $\eta$ in the analysis of the short-time behavior of the WTD.
We show in \Figref{fig:Fig5}, the comparison between $w(\tau)$ and $S(\tau)$, in the case of a weak detection yield $\eta=0.1$, as typically obtained in experiments. 
In contrast to \Figref{fig:Fig3}, both curves get very close one from each other on the same time-window (see \Figref{fig:Fig5}-lower panel).
\bibliography{Biblio_Monte_Carlo}